\documentclass[letterpaper,twocolumn,10pt]{article}
\usepackage{usenix}

\usepackage{tikz}
\usepackage{amsmath}
\usepackage{graphicx}
\usepackage{threeparttable}
\usepackage{subcaption}
\usepackage{tabularx}
\usepackage{array}
\usepackage{amssymb}
\usepackage{booktabs}
\usepackage{url}
\usepackage{balance} 
\usepackage{xspace}
\usepackage{makecell}

\def\revisionDiff{0}

\newcommand{\drevised}[1]{\sout{#1}}

\if\revisionDiff0

    \renewcommand{\drevised}[1]{}
\fi

\newcommand{\subject}[1]{\noindent\textbf{#1}\;}
\newcommand{\subsubject}[1]{\noindent\textit{\underline{#1}}\;}

\newcommand\malurl[1]{\href{notalink}{{\nolinkurl{#1}}}}
\newcounter{finding}
\newcommand{\ignore}[1]{}

\newcommand{\hunter}{\texttt{PHTV Hunter}\xspace}
\newcommand{\arbiter}{\texttt{PHTV Arbiter}\xspace}
\newcommand{\analyzer}{\texttt{PHTV Analyzer}\xspace}
\newcommand{\framework}{\texttt{PHTV-Scout}\xspace}
\newcommand{\phtv}{PHTV\xspace}

\newcounter{insight}
\newcommand{\insight}{
    \refstepcounter{insight}Insight 
    \arabic{insight}:\xspace%
}
\usepackage{filecontents}

\begin{document}

\date{}

\title{\Large \bf When Youth Enter the Algorithmic Wild: Discovering and Understanding Potentially Harmful Teen Videos on Douyin and Kwai}

\author{
{\rm Shaoxuan Zhou}\\
USTC
\and
{\rm Yafei Sun}\\
USTC
\and
{\rm Jing Zhang}\\
USTC
\and
{\rm Xianghang Mi}\\
Monash University
} 

\maketitle

\begin{abstract}

Short-video platforms like Douyin and Kwai have become central to adolescent digital life, but they also risk exposing teens to algorithmically amplified harmful content. Despite its societal importance, the scale, mechanisms, and real-world impact of this exposure remain poorly understood. Measuring it is challenging: recommendation feeds are personalized black boxes, harmful content employs sophisticated evasion tactics, and naive crawlers fail to replicate authentic teen behavior. To bridge this gap, we propose \textbf{PHTV-Scout}, the first large-scale, behaviorally grounded measurement framework for Potentially Harmful Teen Videos (PHTVs). We integrate an offline survey of 683 adolescents with a tri-module online pipeline: (1) \textit{PHTV Hunter} simulates teen accounts to collect recommendation feeds; (2) \textit{PHTV Arbiter}, a LoRA-finetuned multimodal classifier, detects PHTVs with 94.29\% accuracy and 96.41\% precision; and (3) \textit{PHTV Analyzer} performs fine-grained categorization and impact assessment. Over six months, we analyzed 186,727 videos and 51,287 comments, uncovering a troubling 6.11\% PHTV prevalence—dominated by Child Sexual Exploitation Imagery (53.2\%)—and revealing that harmful content thrives through covert interactions (e.g., grooming comments, self-disclosure) and active evasion (semantic camouflage, noise injection). Crucially, while Youth Mode blocks 100\% of PHTVs, its low adoption (30–41\%) leaves most teens unprotected. We further show that exposure is driven not by user identity but by regulation, platform algorithms, and even passive browsing, exposing the fragility of adolescent information environments. Our findings call for a paradigm shift from reactive takedowns to proactive, human-centered safeguards.

\textbf{\textit{Content Warning\;}}This paper contains descriptions of sensitive and potentially harmful content categories related to adolescent safety.

  \end{abstract}
\section{Introduction}
\label{sec:intro}

The global dominance of short-video platforms like TikTok, Douyin, and Kwai has fundamentally reshaped the digital lives of adolescents, offering unprecedented avenues for creativity, socialization, and information consumption~\cite{wang2020humor}. However, this immersive ecosystem also harbors a significant yet understudied risk: the algorithmic amplification of content that is potentially harmful to teenagers. In this paper, we present a large-scale measurement study of \textbf{Potentially Harmful Teen Videos (PHTVs)}—video materials that feature or target adolescents and pose a foreseeable risk of causing behavioral misguidance, psychological distress, or value system distortion. While prior research has extensively examined harmful content in textual or adult-oriented contexts~\cite{young2022much, wang2025detecting, jo2024harmfulyoutubevideodetection}, PHTVs represent a distinct category shaped by adolescent developmental vulnerabilities and the unique mechanics of video-based, algorithmically-curated platforms.  We quantify their prevalence, characterize their distribution mechanisms, and assess their real-world engagement patterns. 

Adolescence is a critical period of neuroplasticity and identity formation, rendering teenagers particularly vulnerable to negative online experiences~\cite{sisk2022stress}. The types of content we identify as PHTVs often glamorize self-injury, promote dangerous challenges, normalize predatory behaviors, or sexualize minors, and prior research has linked adolescent exposure to such material to adverse mental health outcomes and increased risk of behavioral imitation~\cite{liu2017self}. The personalized nature of recommendation feeds means that this exposure is not random but systematically delivered, creating tailored channels for negative influence at scale~\cite{Boeker_2022, baumann2025dynamics}.


Our motivation is further grounded in an offline survey of 683 adolescents (ages 12--17) in China (Section~\ref{sec:teen_user_study}). We find that short-video platforms have become a routine and high-frequency part of teenage life, with substantial daily usage and near-universal interactive behaviors such as liking (54.8--70.1\%), following accounts (38.3--51.3\%), and sharing with peers (45.7--65.6\%). This pattern of heavy, socially participatory engagement aligns with broader trends in adolescent social media use documented by prior studies~\cite{teen_adoption_social_media, adoption_video}. Crucially, we find that while platforms offer technical safeguards such as ``Youth Mode,'' voluntary adoption remains low (30--41\%), primarily because teens perceive it as unnecessary or overly restrictive(Section~\ref{sec:teen_user_study}). At the same time, a majority of respondents report encountering various categories of harmful content, yet a significant proportion show a disconnect between exposure and perceived discomfort. This suggests a dangerous normalization of risk and underscores the urgent need to understand the true landscape of algorithmic exposure of PHTVs.

To address this, our study is guided by three core research questions:
(i) \textit{Negative Social Impact}: What is the prevalence and categorical composition of PHTVs, and how do adolescent users engage with and react to such content?
(ii) \textit{The Arms Race}: How do PHTV creators evade platform moderation, and how effective are platforms’ enforcement mechanisms?
(iii) \textit{Exposure Drivers}: What factors—user attributes, platform algorithms, or regulatory interventions—most significantly influence PHTV exposure in recommendation feeds?

Answering these questions presents formidable challenges. Existing research on harmful content has largely focused on text-based or adult-oriented material~\cite{georgakopoulos2018convolutional, lin2023toxicchat, baldini2022your}, lacking a fine-grained taxonomy and detection framework for adolescent-specific video harms. Methodologically, studying PHTVs is difficult because recommendation feeds are personalized black boxes. Traditional approaches like keyword-based crawling fail to replicate the authentic interaction patterns through which teens organically encounter content, and naive synthetic accounts are easily flagged by platforms, yielding non-representative data. Furthermore, harmful content often employs sophisticated evasion tactics~\cite{yuan2019stealthy,wang2025detecting}, such as semantic camouflage and visual noise injection, making automated detection particularly challenging.

%

To overcome these barriers, we present \framework (Section~\ref{sec:methodology}), a comprehensive, behaviorally grounded measurement framework that integrates empirical insights from our adolescent survey with a multi-stage online analysis pipeline. 
PHTV-Scout comprises three synergistic modules:
\begin{itemize}
    \item \hunter: Mimics authentic adolescent browsing behavior using synthetic accounts grounded in our user study, systematically collecting videos from personalized recommendation feeds across multiple platforms and demographic segments.
    \item \arbiter: A high-precision, multimodal binary classifier built on a LoRA-finetuned Qwen3-VL-8B-Instruct model~\cite{hu2022lora, Qwen-VL}. It fuses video frames, audio transcripts (ASR), on-screen text (OCR), and metadata to filter collected videos, identifying PHTVs with 94.29\% accuracy, 96.41\% precision, and a low false positive rate of only 3.43\%.
    \item \analyzer: Performs fine-grained characterization of identified PHTVs. It conducts multi-class categorization into a novel nine-type taxonomy, longitudinal tracking of video persistence, and in-depth analysis of user comments via sentiment classification and thematic clustering to assess real-world impact.
\end{itemize}
We conducted the entire study under strict ethical protocols, including informed consent, passive data collection, data anonymization, and institutional oversight, ensuring both ecological validity and responsible research practice.


Our six-month measurement campaign, analyzing 186,727 videos and 51,287 comments, yields several critical findings (Section~\ref{sec:findings}):
\begin{itemize}
    \item \textbf{Prevalence}: PHTVs constitute a non-trivial fraction of videos in standard recommendation feeds. We uncover a diverse categorical composition dominated by \textit{Child Sexual Exploitation Imagery}, followed by significant prevalence of \textit{Life-Threatening Risk-Taking Behaviors} and the \textit{Promotion of Unhealthy Lifestyles}.
    \item \textbf{Engagement Paradox}: While PHTVs receive lower overall engagement (likes) than benign content, they foster high-risk micro-interactions: 43\% of comments on sexualized content are positive (often grooming disguised as praise), and 3.5\% of comments on self-harm videos express intent to imitate.
    \item \textbf{Evasion and Enforcement}: 11\% of PHTVs use semantic camouflage (e.g., “dimple art” for self-harm), and 23\% use noise injection to evade detection. Despite platform claims, we observe that enforcement is delayed and selective; up to 96\% of PHTVs remained accessible after one month.
    \item \textbf{Exposure Drivers}: We find that PHTV exposure is driven not by static user attributes (age, gender) but by macro-level regulation (our data shows that a 2025 crackdown reduced prevalence by over 75\%), platform-specific algorithms, and even passive browsing—viewing PHTV comments triggered a 50\% spike in harmful recommendations within 24 hours.
    \item \textbf{Safeguard Paradox}: Youth Mode is technically effective at blocking PHTVs when enabled (as validated by our measurement study), but practically ineffective due to low voluntary adoption among teens (as reported in our user study), leaving most adolescents unprotected.
\end{itemize}


This paper makes three primary contributions:
\begin{enumerate}
    \item We propose \framework, the first comprehensive, behaviorally grounded measurement framework for discovering and characterizing Potentially Harmful Teen Videos on short-video platforms. Its tri-module design (Hunter, Arbiter, Analyzer) enables end-to-end analysis from ecologically valid data collection to fine-grained engagement analysis.
    \item We provide the first comprehensive empirical analysis of the PHTV ecosystem, quantifying its prevalence, mapping its interaction ecology, and exposing the sophisticated creator-platform arms race.
    \item We reveal a critical disconnect in youth online safety: effective technical safeguards are rendered inert by low user adoption and fragile information environments, highlighting the need for human-centered, adaptive protection strategies.
\end{enumerate}


\section{Related Works}
\label{sec:related}
\subject{Harmful Content Detection in Social Media}Social media platforms are central to modern communication, evolving from text-based microblogging to video-centric short-video services. This evolution has been accompanied by the persistent challenge of harmful content, which manifests in various forms and involves an ongoing arms race between platforms and perpetrators~\cite{young2022much, wang2025detecting}. 


\subsubject{Categories of Harmful Online Content} Over the past decade, several categories of harmful content have garnered significant attention from both academia and industry: toxicity and offensive content~\cite{georgakopoulos2018convolutional, baldini2022your, lin2023toxicchat, alcantara-etal-2020-offensive, das2023hatemm}, spam and scams~\cite{wang2010don, gao2012towards, adewole2020twitter}, illicit promotion~\cite{yuan2019stealthy, wang2025detecting}, and misleading content~\cite{boididou2018detection, monti2019fake}. 
Despite being well-studied, these categories often overlap, and their boundaries are subject to ambiguity in annotation criteria and scope. Furthermore, the dynamic interplay between advancements in content moderation techniques and the evolution of social media platforms leads to the emergence of new categories while diminishing the relevance of others.

\subsubject{Detection Approaches} 
Harmful content detection is typically framed as a classification task (binary, multi-class, or multi-label). Over the last decade, detection approaches have evolved from traditional machine learning algorithms (e.g., decision trees, support vector machines)~\cite{kotenko2015evaluation, goyal2016spam}, to low-parameter deep neural networks (e.g., recurrent neural networks, convolutional neural networks)~\cite{georgakopoulos2018convolutional, chandra2018cyberbullying}, and more recently, to fine-tuning pre-trained models (e.g., BERT)~\cite{mozafari2019bert, wang2025detecting}. The rise of large language and vision-language models has introduced a new paradigm via parameter-efficient fine-tuning (PEFT)~\cite{hu2022lora} and training-free in-context learning (ICL)~\cite{brown2020language, lin2024towards, thomas2025supporting}, shifting detection from modality-specific handcrafted features to automatic feature engineering.

\subsubject{Multimodal Harmful Content Detection} Beyond text-only approaches, recent works have begun exploring multimodal detection of harmful content. Multimodal frameworks have been proposed for hate video classification~\cite{prabhu2025comprehensive} and harmful meme detection~\cite{pramanick2021momenta}, leveraging both visual and textual signals. More recently, transformer-based systems have been developed for child-safe content moderation on TikTok~\cite{nguyen2025mtikguard}.



\subsubject{Research Gaps} Prior work exhibits four key limitations. First, harm detection research remains predominantly text-centric~\cite{wang2020humor, adoption_video}. While recent works have begun exploring multimodal detection for specific tasks such as hate videos~\cite{prabhu2025comprehensive}, harmful memes~\cite{pramanick2021momenta}, and general child-safety moderation~\cite{nguyen2025mtikguard}, these efforts target broad harm categories rather than adolescent-specific harms on short-video platforms, and lack a fine-grained taxonomy for nuanced characterization. Second, most studies employ coarse-grained taxonomies~\cite{jo2024harmfulyoutubevideodetection}, lacking the granularity required for nuanced moderation and content characterization. Third, evaluations typically rely on isolated datasets, failing to capture the real-world distribution and prevalence of harmful content. Finally, research predominantly targets adult audiences, neglecting content harmful specifically to minors—a critical oversight given that guardianship assumptions are unfounded~\cite{teen_adoption_social_media}. By 2025, ~60\% of U.S. teens (13–17) are projected to use video-dominant platforms like TikTok and Instagram. Yet youth-mode adoption remains low, and harmful content continues to reach teenagers even when protective mechanisms are enabled~\cite{papadamou2019, tahir2020bring}.While this is a global issue, we focus on China as a representative mature ecosystem; our findings provide early signals for globalizing platforms, and our measurement framework generalizes to other contexts.

\subsubject{Our Contribution} We address these gaps by detecting and analyzing real-world videos that may be harmful to teenagers, leveraging state-of-the-art foundation models. We introduce the first fine-grained taxonomy for harmful teenage videos, capturing realistic risk scenarios on social media. Applying our taxonomy and detection tools, we uncover large-scale, diverse PHTVs and highlight their concerning impact on teenage viewers.



\subject{Foundation Model Adaptation Techniques}
ICL has demonstrated strong performance across various natural language processing (NLP) tasks~\cite{agarwal2024, bertsch2025}. Its accuracy scales from zero-shot to many-shot settings with hundreds of demonstrations~\cite{agarwal2024}, extends to multimodal tasks~\cite{jiang2024}, reduces order-induced variance with more examples~\cite{bertsch2025}, and achieves comparable performance at lower cost via batched inference~\cite{jiang2024} (Full details in Appendix~\ref{appendix:ICL}). 
PEFT techniques, including LoRA, have demonstrated strong performance across various tasks, achieving results comparable to full fine-tuning while being significantly more resource-efficient~\cite{hu2022lora} (Full details in Appendix~\ref{appendix:PEFT}).

 \section{\framework: Automatic Discovery and Attribution of {\phtv}s} 
\label{sec:methodology}

\begin{figure}
    \centering
    \includegraphics[width=0.8\linewidth]{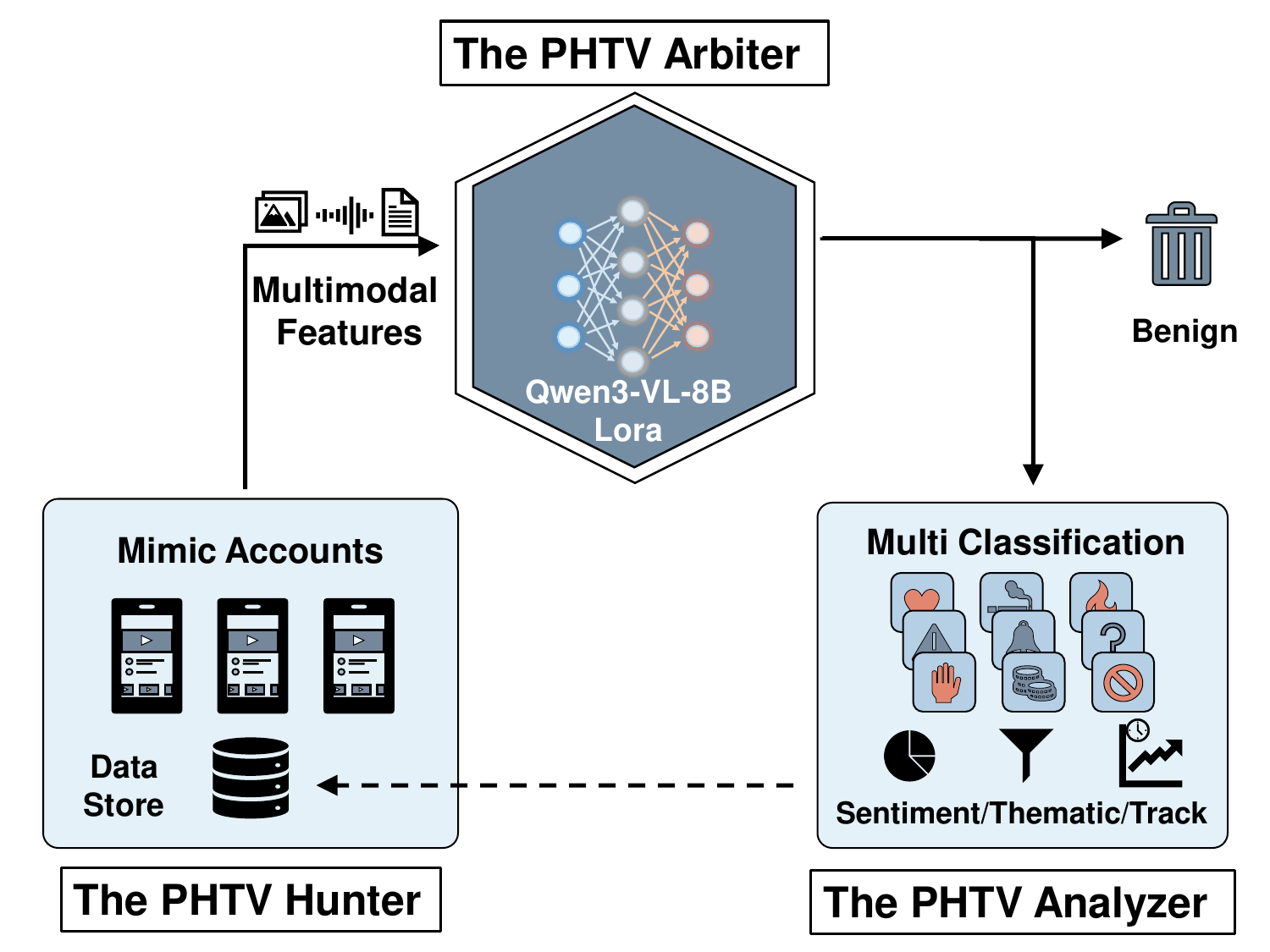}
    \caption{Overview of \framework}
    \label{fig:methodology_overview}
\end{figure}

This section presents our methodology to systematically discover and automatically attribute potentially harmful teen videos (PHTVs) on short video platforms, Our approach is structured into three independent but complimentary  modules: \hunter,  \arbiter, and \analyzer.Figure~\ref{fig:methodology_overview} illustrates the overview of \framework. \hunter is responsible for mimicking realistic teenager interaction and triggering platform recommendation for teen videos including {\phtv}s (Section~\ref{subsec:phtv_hunter}). Given a mix of PHTVs and non-PHTVs discovered by \hunter,  \arbiter (Section~\ref{subsec:phtv_arbiter}) is designed to filter out non-PHTVs and output authentic PHTVs for further analysis. Then, given authentic PHTVs of diverse characteristics, \analyzer (Section~\ref{subsec:phtv_analyzer}) is leveraged to automatically annotate each \phtv with informative attributes (e.g., fine-grained categories), to well facilitate the in-depth characterization presented in Section~\ref{sec:findings}.

\subsection{\hunter: Teenager Account Mimic and Video Collection} \label{subsec:phtv_hunter}

Across discovery and attribution, we adopt ethical practices from previous works~\cite{wang2025detecting,jo2024harmfulyoutubevideodetection} and carefully design our methodology to avoid potential ethical risks, as concluded in Appendix~\ref{appendix:ethical_considerations}.

The \hunter module represents the first phase of our pipeline, designed to achieve two primary objectives. First, it systematically discovers and collects short videos that a typical teenager might encounter but are less likely to be recommended to adults, with a particular focus on potentially harmful teen videos (PHTVs). These collected videos form the foundational dataset for subsequent detection and attribution modules. Second, once a PHTV is identified, \hunter periodically revisits the video to collect its comments and monitor its temporal evolution, enabling us to longitudinally analyze its engagement and persistence.

\subject{Challenges}
\label{sss:challenges_video_discovery}
The design of \hunter faced two major challenges. The first was the heterogeneity of video access mechanisms across platforms. Platforms like Kwai and Douyin employ distinct recommendation algorithms and content delivery systems, requiring platform-specific crawling strategies to ensure consistent and reliable data collection. The second challenge was triggering videos that a typical teenager might be exposed to, which necessitated simulating authentic adolescent user behaviors. This required careful consideration of user interaction patterns, including likes, follows, and browsing habits, to replicate the pathways through which teenagers organically encounter content.

\subject{Insights from User Study and Prior Research}
\label{sss:inspirations_user_study}
The design of \hunter was informed by two key sources of inspiration. First, an offline user behavior study involving 683 adolescents aged 12–17 provided empirical insights into platform usage, daily engagement, and exposure to harmful content, shows that interaction is near-universal: liking (54.8–70.1\%), following accounts (38.3–51.3\%), and sharing with peers (45.7–65.6\%) (See more details in Section~\ref{sec:teen_user_study}). This user study highlighted the importance of simulating realistic teenage interactions to ensure ecological validity. Second, prior research on recommendation amplification mechanisms in short video platforms~\cite{Boeker_2022,baumann2025dynamics} revealed how user signals, such as watch behavior and likes, can rapidly steer recommendation algorithms into narrow interest areas. These findings underscored the need for behaviorally grounded synthetic accounts to mimic the dynamics of adolescent content exposure.


\subject{Key Design Details}
\label{sss:synthetic_account_design}
To trigger adolescent-targeted recommendations, we simulate teenager accounts that passively collect daily feeds. As summarized in Table~\ref{tab:account_design}, two account types are deployed: (1) Cold-start accounts represent new users with no interaction history, configured at ages 13, 15, and 17 to capture default algorithmic behavior; and (2) Behavioral-seed accounts mimic light engagement by initially liking 10–20 common videos (e.g., campus life) and following 30–50 typical creators (e.g., study tips), reflecting early exploration patterns observed in our user study to probe how initial interactions shape recommendations.

\begin{table}
\centering
\caption{Synthetic account configuration based on user study.}
\label{tab:account_design}
\begin{tabular}{lcc}
\toprule
Account Type & Account  & Age Setting \\
\midrule
\multicolumn{3}{c}{\textit{No likes/follows/searches}} \\
Cold-start & A ($n=1$)      & 13 (represents 12--13)  \\
          & B ($n=1$)       & 15 (represents 14--15)  \\
          & C ($n=1$)       & 17 (represents 16--17)  \\
\addlinespace[0.5em]
\midrule
\multicolumn{3}{c}{\textit{ Like 10--20 videos; follow 30--50 accounts}} \\
Behavioral-seed 
          & D ($n=3$)        & 13  (represents 12--13)   \\
          & E ($n=3$)        & 15  (represents 14--15)   \\
          & F ($n=3$)        & 17  (represents 16--17)    \\
\bottomrule
\end{tabular}
\end{table}
\subject{Key Implementation Details}
\label{sss:crawling_strategy} When it comes to account setup, 
due to platform regulations requiring identity verification for users under 14, the profile age of low-age accounts is set to 14 years as a close approximation for representing 12–13-year-old adolescents, while other groups are set up as normal. 

All crawled videos are stored on dedicated research servers along with necessary metadata including video ID, publish timestamp, and engagement metrics (likes). Also, no interactive actions (e.g., liking, commenting) are performed during crawling to ensure passive data collection and maintain ecological validity. More details about the video collection are provided in Appendix~\ref{appendix:video_collection}.


\subsection{\arbiter: Feature Engineering and Binary Classification}
\label{subsec:phtv_arbiter}
The videos collected by \hunter comprise a mix of both PHTVs and non-PHTVs. To address this, we design \arbiter as a filtering mechanism to identify and retain only PHTVs. This task is formulated as a binary classification problem, where the goal is to determine whether a given video falls into the general category of potentially harmful teen videos. 

Unlike prior studies on harmful online content, which often rely solely on either textual or visual features, \arbiter leverages a combination of both modalities. This approach is further enhanced by utilizing the latest advancements in foundation models and fine-tuning techniques. When selecting the optimal classification setup, we prioritize configurations that achieve a low false positive rate and high precision. This ensures that the identified PHTVs are of high fidelity, which is critical for maintaining the credibility of the subsequent measurement results based on these videos.

Below, we outline the key design decisions and present the evaluation results of \arbiter.

\subject{Ground Truth Annotation}  
We manually annotated 3,510 videos to construct our ground truth dataset—1,755 harmful (all involving adolescents as subjects or targets) and 1,755 benign—ensuring balance and alignment with our study focus. Harmful videos span eight subcategories (Table~\ref{tab:binary_dataset_composition}),  ensuring robust coverage of the PHTV taxonomy.

We constructed the dataset via a snowball-based seeding process: initial PHTVs discovered through passive recommendation browsing yielded recurring keywords, which were iteratively expanded until saturation. Simulated teenage accounts seeded with these keywords and author interactions triggered personalized recommendation feeds for harvesting. This approach better reflects how adolescents organically encounter harmful content than static keyword crawling. Full details are provided in Appendix~\ref{appendix:ground_truth}.


We labeled all videos following the annotation protocol detailed in Section~\ref{sss:multi_classification}; the dataset was split 90\%/10\% for training/validation and testing, respectively.

\begin{table}
\centering
\caption{Composition of the annotated PHTV/non-PHTV dataset (balanced 50/50 split). All eight subcategories (Detail definitions in Table~\ref{tab:taxonomy}) are well represented.}
\label{tab:binary_dataset_composition}
\begin{tabularx}{\linewidth}{@{}>{\raggedright\arraybackslash}Xc@{}}
\toprule
Category & Percentage \\
\midrule
C1: Aberrant Social Media Challenges & 5.5\% \\
C2: Child Sexual Exploitation Imagery & 34.1\% \\
C3: Glorification of Youth Violence & 4.0\% \\
C4: Intentional Self-Injury \& Extreme Body Modifications & 7.4\% \\
C5: Life-Threatening Risk-Taking Behaviors & 11.1\% \\
C6: Maladaptive Behavioral Influence & 6.9\% \\
C7: Predatory Recruitment of Minors & 18.3\% \\
C8: Promotion of Unhealthy Lifestyles & 12.7\% \\
\midrule
Total Harmful Videos & 50.0\% \\
Total Benign Videos & 50.0\% \\
\bottomrule
\end{tabularx}
\end{table}

\subject{Classification Algorithms} Foundation models have achieved state-of-the-art performance in both NLP and vision tasks. Therefore, we focus our classification algorithm exploration on two paradigms built upon foundation models: in-context learning (ICL) and LoRA, a parameter-efficient fine-tuning technique. Both paradigms support automatic feature engineering, allowing us to concentrate on designing the raw input representations. Based on PHTV case studies, we observe that harm cues may exist in various modalities, including audio, text (e.g., post titles), text embedded in video frames, or purely visual elements. To address this, we design a multimodal input representation, as detailed below.

\subject{Multimodal Input Representation} Given a PHTV $v$ and its associated metadata (e.g., video title), the video is encoded into a fused multimodal input representation:
\begin{equation}
    \mathbf{x}_v = \left( t_v,\, a_v,\, o_v,\, \{f_i\}_{i=1}^{N} \right),
\end{equation}
where $t_v$ represents the title, $a_v$ the audio transcript extracted by automatic speech recognition (ASR), $o_v$ the aggregated text extracted from the video via optical character recognition (OCR), and $\{f_i\}_{i=1}^{N}$ a sequence of video frames sampled at a configurable frame rate $r_f$ (default: $r_f = 0.5$ FPS). The total number of frames $N$ is calculated as $N = \lfloor T_v \cdot r_f \rfloor$, where $T_v$ is the video duration in seconds. This multimodal representation is directly fed into the classification pipeline.

To ensure comprehensive content understanding, we extract linguistic information from both the audio and visual modalities of each video. This step is critical for detecting harmful content that may be obfuscated in speech or text overlays. More details about ASR and OCR are provided in Appendix~\ref{appendix:asr_ocr} 





\subject{Evaluation} 
\label{sss:binary_classification}
We evaluated binary classification using in-context learning (ICL) and LoRA fine-tuning via grid search across model architectures, prompts, shots, input compositions (ICL), and rank/scaling hyperparameters (LoRA). Performance was measured by accuracy, precision, recall, F1-score, and FPR on both held-out ground truth data and an out-of-distribution test set of 1,000 videos from live streams to assess real-world generalization.

Evaluation showed that ICL-based methods, even with chain-of-thought prompting, struggled to exceed 80\% accuracy and often had recall rates below 50\%. In contrast, LoRA fine-tuned \texttt{Qwen3-VL-8B-Instruct} achieved 94.29\% accuracy, 96.41\% precision, and 92.00\% recall on the internal test set, demonstrating strong generalization on the OOD dataset (96.80\% accuracy, 95.60\% precision, 97.95\% recall, 4.30\% FPR). Summarized results appear in Table~\ref{tab:all_model_comparison_main}, and the prompt template used in the best-performing setup is detailed in Appendix~\ref{appendix:prompt}.

An ablation study further revealed that analyzing only the first 10 seconds of a video achieved 93.4\% accuracy, suggesting a promising path to reduce computational costs while maintaining strong detection performance. Given the superior performance of the LoRA fine-tuned Qwen3-VL-8B-Instruct, we selected it as the final binary classifier. Additional details, including full model comparisons and deployment configurations, are provided in Appendix~\ref{appendix:performance_and_eval}.

\begin{table}[t]
    \centering
    \caption{Performance comparison of candidate binary classifiers. All ICL variants use the same structured prompt with full category definitions (see Appendix~\ref{appendix:prompt}). The best-performing LoRA model is highlighted in bold.}
    \label{tab:all_model_comparison_main}
    \begin{threeparttable}
    \small
        \begin{tabular}{@{}lccccc@{}}
            \toprule
            Model & \makecell{Acc. \\ (\%)} & \makecell{Prec. \\ (\%)} & \makecell{Rec. \\ (\%)} & \makecell{F1 \\ (\%)} & \makecell{FPR \\ (\%)} \\
            \midrule
            \multicolumn{6}{c}{\textit{Fine-tuned Pre-trained Models}} \\
            Early Fusion (Text+Video)          & 92.57 & 94.08 & 90.86 & 92.44 & 5.71 \\
            Late Fusion (Text+Video)           & 77.43 & 86.92 & 64.57 & 74.10 & 9.71 \\
            VideoMAE (Video-only)              & 72.29 & 71.91 & 73.14 & 72.52 & 28.57 \\
            BERT (Text-only)                   & 78.57 & 76.88 & 81.71 & 79.22 & 24.57 \\
            \midrule
            \multicolumn{6}{c}{\textit{In-Context Learning (ICL)}} \\
            Gemma3-12B                         & 53.43 & 87.50 & 8.00  & 14.66 & 1.14 \\
            \makecell[l]{Qwen-VL-MAX \\ (128-shot)}       & 79.66 & 82.10 & 76.00 & 78.93 & 16.67 \\
            \makecell[l]{GPT-4-Turbo \\ (CoT-EN)\tnote{a}}        & 54.29 & 66.67 & 17.14 & 27.27 & 8.57 \\
            \makecell[l]{Claude-Opus-4.1 \\ (CoT-EN)\tnote{a}}     & 63.87 & 89.83 & 30.81 & 45.89 & 3.45 \\
            \midrule
            \multicolumn{6}{c}{\textit{Fine-tuned Models}} \\
            \makecell[l]{\textbf{Qwen3-VL-8B} \\ \textbf{(LoRA)}} & \textbf{94.29} & \textbf{96.41} & \textbf{92.00} & \textbf{94.15} & \textbf{3.43} \\
            \bottomrule
        \end{tabular}
        \begin{tablenotes}
            \item [a] CoT-EN refers to Chain-of-Thought prompting in English.
        \end{tablenotes}
    \end{threeparttable}
\end{table}

\subsection{\analyzer: Automatic Annotation and Engagement Analysis} 
\label{subsec:phtv_analyzer}
Given the PHTVs detected by \arbiter, \analyzer \textit{automatically} annotates each PHTV with informative attributes that are critical for understanding the nature and reception of PHTVs. Similar to \arbiter, this is achieved through the adoption of carefully selected machine learning techniques, as elaborated below.

Currently, \analyzer is designed to support the annotation of three groups of attributes: fine-grained PHTV categories, sentiment attributes, and thematic attributes for user comments associated with each PHTV. Below, we briefly describe how automatic annotation for these attributes is achieved.

\subject{Fine-grained Multi-class Categorization of PHTVs} \label{sss:multi_classification}
While the binary classifier in \arbiter effectively identifies PHTVs, binary labels alone cannot explain harm mechanisms or prioritize mitigation efforts. To enable fine-grained analysis, we developed a multi-class classifier that automatically annotates each PHTV into a taxonomy of nine well-defined harm categories (Table~\ref{tab:taxonomy}).

\begin{table*}[t]
\centering
\caption{Nine-category taxonomy for fine-grained categorization of PHTVs.}
\label{tab:taxonomy}
\begin{tabular}{@{}>{\raggedright\arraybackslash}p{0.30\textwidth}>{\raggedright\arraybackslash}p{0.67\textwidth}@{}}
\toprule
\textbf{Category Name} & \textbf{Definition} \\
\midrule
\textbf{Aberrant Social Media Challenges} & Content depicting bizarre, attention-seeking behaviors that lack inherent physical danger but deviate significantly from normative social conduct (e.g., mimicking animals eating off the floor, filtering bubble tea through socks, inserting spicy snack strips into nostrils). \\
\addlinespace[0.5em]

\textbf{Child Sexual Exploitation Imagery (CSEI)} & Content featuring minors engaging in sexualized performances or exhibiting sexually suggestive behaviors and explicit nudity (e.g., provocative dancing, sexually suggestive poses). \\
\addlinespace[0.5em]

\textbf{Glorification of Youth Violence} & Content that normalizes, romanticizes, or celebrates violent behavior and aggressive conduct among minors (e.g., staged physical fights, documented bullying incidents, verbal abuse and threats). \\
\addlinespace[0.5em]

\textbf{Intentional Self-Injury \& Extreme Body Modifications} & Content depicting or glorifying deliberate physical self-harm or unsafe bodily alterations performed by minors (e.g., cutting or burning skin, unsafe DIY piercing procedures). \\
\addlinespace[0.5em]

\textbf{Life-Threatening Risk-Taking Behaviors} & Content showcasing activities with substantial potential for severe physical injury or fatality among youth (e.g., fainting challenges, dangerous motorcycle stunts, fire-breathing performances). \\
\addlinespace[0.5em]

\textbf{Maladaptive Behavioral Influence} & Content promoting behaviors that undermine adolescents' healthy development or social functioning (e.g., encouraging school dropout, inciting vandalism, promoting excessive food waste). \\
\addlinespace[0.5em]

\textbf{Predatory Recruitment of Minors} & Content involving the solicitation or manipulation of minors for financial exploitation or illicit activities (e.g., recruiting underage labor, coercing financial transfers, inducing participation in fraud schemes). \\
\addlinespace[0.5em]

\textbf{Promotion of Unhealthy Lifestyles} & Content that romanticizes or encourages harmful health practices and health-compromising behaviors among youth (e.g., smoking cigarettes, alcohol consumption, menstruation rejection). \\
\addlinespace[0.5em]

\textbf{Other Digital Harm to Minors} & Residual category encompassing harmful content directed at minors that does not align with the preceding categories. \\
\bottomrule
\end{tabular}
\end{table*}

\noindent\textbf{Taxonomy Development and Annotation Protocol.}
The nine-category taxonomy was grounded in China's \textit{Regulations on the Protection of Minors on the Internet} and refined through iterative review of collected PHTVs. An initial codebook was tested on a 200-video pilot study by three trained annotators, who resolved disagreements via discussion and updated category definitions based on edge cases. The refined taxonomy was then applied to annotate all 1,755 PHTVs, with ongoing weekly calibration meetings maintaining consistency and achieving pairwise Cohen's $\kappa > 0.90$. Data collection followed a saturation criterion: across five annotation rounds of $\sim$2,000 videos each ($\sim$10,000 total), no new harmful subcategories emerged in the final two rounds, at which point we concluded the taxonomy was comprehensive and ceased further collection.

Key decision boundaries between frequently confused categories are: C1 (socially deviant but non-dangerous) vs.\ C5 (substantial injury/death risk); C4 (deliberate self-harm as the goal) vs.\ C5 (harm as a byproduct of risky activity); C6 (undermines social functioning) vs.\ C8 (health-compromising behaviors). Full protocol and annotator training details are provided in Appendix~\ref{appendix:ground_truth}.



We trained the multi-class classifier using the same architectural and training paradigm as the best-performing binary model, but with a modified prompt template tailored for the multi-class classification task (Appendix~\ref{appendix:multi_model_configuration}).

This LoRA-based model achieved strong performance, with a weighted precision of 93.91\%, a weighted recall of 93.99\%, a macro-averaged precision of 91.25\%, and a macro-averaged recall of 89.88\%. Notably, this performance is comparable to or slightly better than in-context learning (ICL) using a much larger state-of-the-art model (e.g., \textit{Qwen-VL-MAX\cite{Qwen-VL}}). Considering its smaller model size and significantly higher inference efficiency, this result is particularly impressive. All technical details are provided in Appendix~\ref{appendix:multi_classifier_details}.

\subject{Longitudinal PHTV Tracking} \label{sss:comment_tracking_new}
Upon identifying a video as PHTV, we apply two periodic monitoring processes: (1) crawling engagement metrics (e.g., comments, likes) to assess engagement dynamics after annotation, and (2) revisiting the video over time to track its persistence and the evolving arms race between content moderation and PHTV distribution. Both processes leverage \hunter for daily jobs while respecting platform rate limits to avoid server pressure (see Appendix~\ref{appendix:longtitudinal_tracking} for implementation details). Together, they transform static classification into dynamic monitoring, enabling a comprehensive understanding of how PHTVs spread, persist, and engage adolescent users.




\subject{Sentiment and Thematic Analysis of User Comments} \label{sss:sentiment_analysis_new}
To assess the real-world impact of PHTVs, we analyze user comments collected through longitudinal tracking, revealing how adolescents perceive and react to harmful content within their personalized feeds. Our analysis comprises two components: (1) three-class sentiment classification to profile comment-level attitudes toward PHTVs, and (2) thematic analysis to identify recurring patterns and high-risk behaviors such as imitation or glorification of harmful acts.

\subsubject{Sentiment Analysis} We employ \texttt{Qwen3-Max-Instruct}~\cite{qwen3max} as a zero-shot classifier to categorize comments into \textit{positive} (admiration/support), \textit{negative} (criticism/concern), or \textit{neutral} (factual statements/@-mentions). Evaluated on 1,000 manually labeled comments (183 positive, 179 negative, 638 neutral), the model achieves 98.70\% weighted precision and recall (macro precision: 98.91\%, recall: 98.13\%), confirming its reliability for large-scale deployment.

\subsubject{Thematic Analysis} We employ a human-in-the-loop approach: \texttt{Qwen3-Max-Instruct} performs semantic clustering on comments collected from all PHTV categories, generating initial thematic labels that two annotators with expertise in youth digital culture iteratively refine. This process yields 10 consolidated thematic categories (Table~\ref{tab:thematic_categories}).

\begin{table*}
\centering
\caption{Thematic categories (TC1–TC10) identified in PHTV comment analysis (More details see Appendix~\ref{appendix:sentiment_thematic}).}
\label{tab:thematic_categories}
\begin{tabular}{@{}ll@{}}
\toprule
\textbf{Category} & \textbf{Definition} \\
\midrule
TC1: Supportive Feedback & Encouragement or validation of creators. \\
TC2: Critical Feedback & Skepticism about authenticity or ethics. \\
TC3: Hostile Language & Abusive, threatening, or discriminatory remarks. \\
TC4: Imitative Behavior Expression & Explicit intent to replicate video behaviors. \\
TC5: Social Summoning (@xxx) & Use of ‘@’ to refer to other users. \\
TC6: Resonance-Driven Self-Disclosure & Trauma/experience-based personal sharing triggered by emotional resonance. \\
TC7: Predatory Recruitment (Minors) & Solicitation for risky/illegal activities targeting minors. \\
TC8: Risky Lifestyle Promotion & Promotion of e-cigarettes, tobacco, or similar substances. \\
TC9: Sexual Harassment (Minors) & Sexualized comments or contact requests toward young creators. \\
TC10: Other / Noise & Uninterpretable, off-topic, or nonsensical content. \\
\bottomrule
\end{tabular}
\end{table*}
 
Integrating sentiment and thematic results with video categories and engagement metrics reveals how PHTVs are received and amplified across the platform ecosystem. Full methodological details of sentiment analysis and thematic taxonomy are in Appendix~\ref{appendix:sentiment_thematic}.

\section{In-Depth Characterization of PHTVs} 
\label{sec:findings}

Based on the methodologies outlined in Section~\ref{sec:methodology}, we have identified a substantial and diverse corpus of PHTVs, enriched with detailed annotations. This dataset enables us to conduct a comprehensive analysis of PHTVs, focusing on three critical research questions. First, we investigate the negative impact that PHTVs exert on adolescent social media users (Section~\ref{subsec:negative_impact}). Second, we examine the ongoing arms race between platforms and PHTV creators, exploring the strategies employed by both sides and identifying key lessons that can inform future mitigation efforts (Section~\ref{subsec:moderation_evasion}). Finally, we analyze the factors that influence the accessibility and reach of PHTVs, with particular emphasis on their exposure to adolescent users within social media ecosystems (Section~\ref{subsec:exposure_factors}).

\subsection{Negative Social Impact of PHTVs} \label{subsec:negative_impact}







Understanding the negative social impact of PHTVs is not only an academic question—it has direct implications for adolescent safety, platform accountability, and regulatory design. If we can systematically characterize how PHTVs manifest and how users respond to them, stakeholders gain actionable insights: platforms can refine detection heuristics beyond surface-level signals, policymakers can prioritize high-risk content categories for intervention, and educators can develop targeted media literacy programs to inoculate vulnerable youth. Without such evidence, mitigation efforts risk targeting the wrong harms or overlooking covert but high-consequence interactions. 

Ideally, to fully understand the impact of PHTVs, one would conduct a longitudinal study that continuously tracks a cohort of adolescents—monitoring their exposure to harmful content, behavioral responses, and psychological trajectories over time. Such an approach could establish causal links between PHTV consumption and real-world outcomes. However, this is ethically and practically infeasible: exposing minors to known harmful content violates research ethics, and long-term behavioral monitoring at scale raises significant privacy and consent challenges.

Instead of direct observation, we adopt a complementary measurement strategy: by analyzing the \textit{content landscape} and the \textit{user interaction ecology} surrounding PHTVs, we infer their potential for negative social impact. Specifically, we examine (1)what the scale and categorical composition of PHTVs organically recommended to adolescent-like accounts are, and (2) how users actually engage with such content—through likes, comments, and peer discourse. This approach reveals the structural conditions under which harm may propagate: the magnitude and types of content being delivered, the affective responses they elicit, and the behavioral signals they inspire.

Specifically, to answer this question, we rely on the following measurement points:
\begin{itemize}
    \item \textbf{PHTV prevalence and categorical distribution}: measured over 168,165 videos collected from standard recommendation feeds on Douyin and Kwai (Jul., Oct. and Dec. 2025) using simulated adolescent accounts.
    \item \textbf{Aggregate user engagement}: quantified via likes on PHTVs versus benign videos.
    \item \textbf{Sentiment distribution in user comments}: classified as positive, neutral, or negative across all PHTV categories.
    \item \textbf{Thematic patterns in user comments}: measured by grouping user comments into a high-coverage thematic taxonomy consisting of our ten categories(Table~\ref{tab:thematic_categories}).
\end{itemize}

\medskip
\noindent \textbf{\insight The PHTV landscape is structurally dominated by the sexualized exploitation of minors, while high-imitation-risk behaviors constitute a substantial secondary layer of harm. } 

This is first distilled from the categorical distribution of 10,268 PHTVs identified in our corpus of 168,165 videos. As shown in Table~\ref{tab:phtv_distribution_overall}, Child Sexual Exploitation Imagery (C2) alone accounts for \textbf{53.2\%} of all harmful content—featuring minors in suggestive poses, revealing attire, or choreographed dances, often accompanied by benign textual framing such as “Normal dressing style with no negative influence.”  



Furthermore, Life-Threatening Risk-Taking Behaviors (C5, 16.8\%) and Promotion of Unhealthy Lifestyles (C8, 10.4\%) together represent \textbf{27.2\%} of PHTVs, forming a significant secondary tier.  These contents either include high-risk activities such as human pyramids, fire-breathing, wheelie riding, making homemade explosives, or promote harmful behaviors such as smoking (including eating cigarette butts), drinking, chewing betel nuts, resisting menstruation (“to hell with my period”)—all of which are highly visually appealing and easy to imitate.

Finally, acute but low-frequency threats persist at non-negligible levels: categories such as Aberrant Social Media Challenges (C1, 2.5\%), Maladaptive Behavioral Influence (C6, 2.7\%), and Predatory Recruitment of Minors (C7, 2.8\%) consistently appear in recommendations despite their rarity, with C7 often involving direct solicitations like “What grade are you in? I’ll send you money.” Concrete examples for all categories are provided in Appendix~\ref{appendix:phtv_examples}.

\begin{table}
\centering
\caption{Overall distribution of PHTVs across harm categories and account types (all platforms, July--December 2025).}
\label{tab:phtv_distribution_overall}
\begin{threeparttable}
\begin{tabularx}{\linewidth}{@{}>{\raggedright\arraybackslash}Xrc@{}}
\toprule
Category & Count & \%\tnote{1} \\
\midrule
C1: Aberrant Social Media Challenges & 254 & 2.5\% \\
C2: Child Sexual Exploitation Imagery & 5,468 & 53.2\% \\
C3: Glorification of Youth Violence & 799 & 7.8\% \\
C4: Intentional Self-Injury \& Extreme Body Modifications & 386 & 3.8\% \\
C5: Life-Threatening Risk-Taking Behaviors & 1,724 & 16.8\% \\
C6: Maladaptive Behavioral Influence & 281 & 2.7\% \\
C7: Predatory Recruitment of Minors & 283 & 2.8\% \\
C8: Promotion of Unhealthy Lifestyles & 1,073 & 10.4\% \\
\midrule
Total PHTVs & 10,268 & 6.11\% \\
Total Videos Collected & 168,165 & 100.0\% \\
\bottomrule
\end{tabularx}
\begin{tablenotes}
    \item [1] Percentages for harm categories are calculated over all PHTVs, while the percentage for `Total PHTVs' is calculated over `Total Videos Collected'.
\end{tablenotes}
\end{threeparttable}
\end{table}


\medskip
\noindent
\textbf{\insight Despite receiving lower engagement than benign videos, PHTVs involve high-stakes micro interactions that disproportionately amplify harm within specific sub categories.}  

This is first reflected in reduced engagement on harmful content, quantified by the Relative Engagement Intensity (REI), defined as:
\begin{equation}
    \mathrm{REI} = \frac{\overline{E}_{\text{Harmful}}}{\overline{E}_{\text{Benign}}},
\end{equation}
where $\overline{E}$ denotes average likes per video. Based on 500 randomly sampled PHTVs and 500 benign videos per platform, we calculated the REI and found it consistently below 1—specifically, \textbf{0.50 }on Douyin and \textbf{0.89} on Kwai—indicating that PHTVs receive significantly fewer likes than benign content on both platforms.


However, this macro-level trend masks profound heterogeneity in user responses—revealed through two complementary lenses: sentiment analysis and thematic analysis of 51,287 comments on PHTVs.

\begin{figure}
\centering
\begin{subfigure}[b]{0.48\linewidth}
\centering
\includegraphics[width=\linewidth]{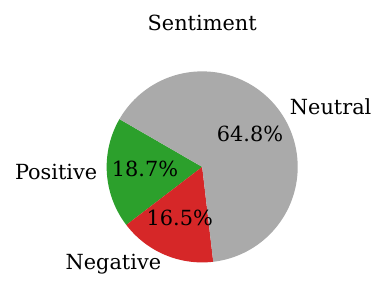}
\caption{Overall sentiment distribution}
\label{fig:sent_pie}
\end{subfigure}
\hfill
\begin{subfigure}[b]{0.48\linewidth}
\centering
\includegraphics[width=\linewidth]{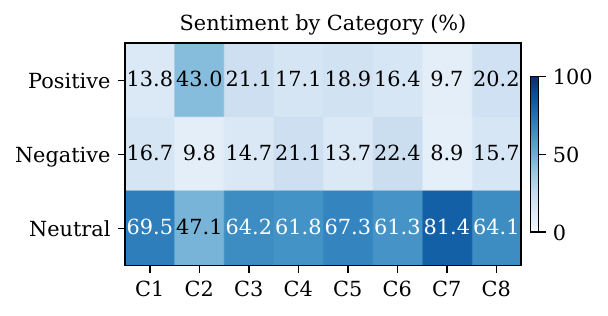}
\caption{Sentiment by PHTV category}
\label{fig:sent_heatmap}
\end{subfigure}
\caption{Sentiment analysis of PHTV comments (N=51,287).}
\label{fig:sentiment_combined}
\end{figure}

From the \textit{sentiment analysis}, we find that while the majority of comments are neutral (64.82\%, primarily driven by social mentions like ``@xxx'' and factual queries), a substantial affective undercurrent persists. Notably, \textbf{C2 (Child Sexual Exploitation Imagery) exhibits the highest positivity rate across all categories at 43.01\%} (Figure~\ref{fig:sent_heatmap}). These positive expressions often take the form of seemingly innocent praise—such as ``so cute!'' or ``you’re adorable!''—but, when contextualized with the visual content featuring minors in sexualized poses, likely constitute \textit{predatory grooming language}. In contrast, \textbf{C4 (Intentional Self-Injury)} and \textbf{C6 (Maladaptive Behavioral Influence)} show the highest negativity rates (21.14\% and 22.35\%, respectively), reflecting strong community condemnation of self-harm and dysfunctional behaviors.

\begin{figure}
\centering
\begin{subfigure}[b]{0.48\linewidth}
\centering
\includegraphics[width=\linewidth]{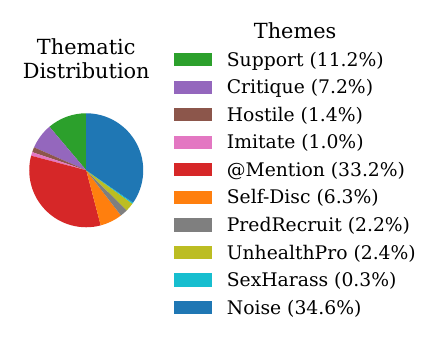}
\caption{Overall thematic distribution}
\label{fig:theme_pie}
\end{subfigure}
\hfill
\begin{subfigure}[b]{0.48\linewidth}
\centering
\includegraphics[width=\linewidth]{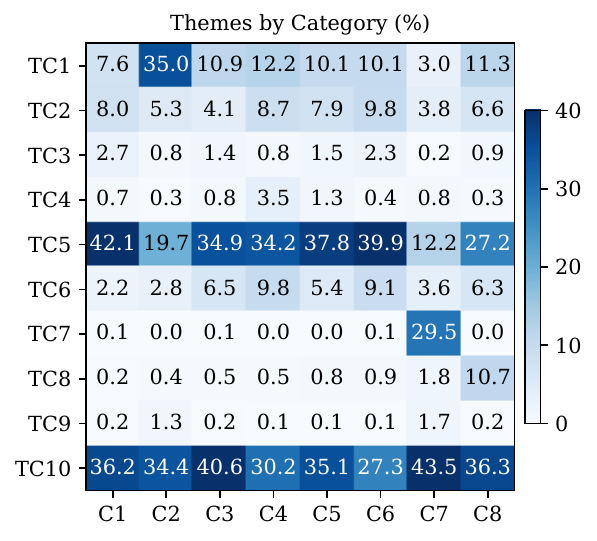}
\caption{Thematic by PHTV category}
\label{fig:theme_heatmap}
\end{subfigure}
\caption{Thematic analysis of PHTV comments (N=51,287).}
\label{fig:theme_combined}
\end{figure}

Crucially, sentiment alone cannot capture the full risk landscape. The \textit{thematic analysis} uncovers high-severity semantic patterns that are invisible through sentiment classification(See Table~\ref{tab:thematic_categories}). For instance:
\begin{itemize}
    \item For \textbf{C7 (Predatory Recruitment of Minors)}, 29.52\% of comments fall under the theme TC7(Predatory Recruitment Targeting Minors), featuring direct, transactional solicitations such as ``Anyone want to be a punching bag? 288 RMB/hour''—often using coded language to bypass automated moderation.
    \item In \textbf{C8 (Promotion of Unhealthy Lifestyles)}, 10.67\% of comments explicitly engage in TC8(Promotion of Risky Lifestyles), normalizing e-cigarette use, extreme dieting, or unregulated supplements as peer-endorsed choices.
    \item Most alarmingly, among comments under \textbf{C4 (Intentional Self Injury \&Extreme Body Modiﬁcations)}, 3.50\% show TC4(Behavioral Imitation Expression), where users explicitly endorse replication (e.g., ``I want to drill a few holes in my face [cool girl]''), expressing an intent to replicate these behaviors in their own lives.
\end{itemize}
These patterns distributed differently in different categories and are visualized in Figure~\ref{fig:theme_heatmap}.


\medskip
\noindent
\textbf{\insight PHTVs can trigger resonance-driven self-disclosure and social amplification—even through seemingly benign interactions—suggesting a potential mechanism for emotional contagion and content amplification.} 

This is distilled from the thematic analysis of comments, which reveals that \textbf{9.82\% of comments under C4 (Intentional Self-Injury)} and \textbf{9.09\% under C6 (Maladaptive Behavioral Influence)} fall into the TC6(Resonance-Driven Self-Disclosure) category. These users share personal narratives of past self-harm, or emotional crises explicitly triggered by the video content (e.g., ``I also like wearing three lip piercings and a dozen ear piercings, and I even had a Dracula piercing [laughing emoji] because my parents didn't agree, I took them out and they all healed up''). While such disclosures may stem from a genuine need for identification or support, they often receive peer validation like ``you’re not alone'' or ``same here,'' 
creating feedback loops where distress may become normalized as a form of identity or social currency. In the absence of professional intervention, there is a concern that such exchanges could inadvertently reinforce maladaptive coping mechanisms.

Moreover, our analysis uncovers a pervasive yet overlooked vector of harm: \textbf{TC5(Social Summoning)}. Across all PHTV categories, \textbf{33.20\% of comments consist of ``@xxx'' mentions} (Figure~\ref{fig:theme_pie}), often without additional text. While these appear neutral—and are frequently filtered out as noise—they may serve to propagate harmful content to peers. For instance, a comment like ``@xxx'' under a C5 video showing dangerous riding explosives may function as an implicit invitation to view or imitate. Similarly, tagging friends under C2 or C8 videos can normalize sexualized or unhealthy behaviors within adolescent social circles. This form of passive sharing bypasses platform moderation (as it lacks explicit keywords) yet effectively extends the audience of PHTVs beyond algorithmic recommendations.

Together, these patterns reveal a dual amplification mechanism: (1) \textit{internal}, where vulnerable users internalize and re-express harm through self-disclosure; and (2) \textit{external}, where users inadvertently—or intentionally—spread harmful content through social mentions. Both operate beneath the surface of aggregate metrics, yet collectively extend the reach and engagement intensity of PHTVs.



\subsection{How PHTV Creators Outmaneuver Platform Moderation}
\label{subsec:moderation_evasion}

Short-video platforms assert that harmful content targeting minors is swiftly detected and removed. Yet, as shown in Section~\ref{subsec:negative_impact}, PHTVs not only reach adolescent users but also sustain high-risk interactions—implying a failure in timely enforcement. To understand this gap, we frame the problem as an \textit{asymmetric arms race}: creators deploy low-cost evasion tactics to bypass automated detectors, while platforms struggle with delayed and incomplete removal. Analyzing this contest is critical—it exposes the limitations of current moderation, reveals why harmful content persists despite platform claims, and informs the design of more resilient detection systems.

To characterize this arms race, we rely on the following measurement points:
\begin{itemize}
    \item \textbf{Creator-side evasion tactics}, measured through the manual annotation of 1,755 representative PHTVs by human reseachers. 
    \item \textbf{Strategic upload timing}, derived from 10,000 randomly sampled PHTVs, analyzed by hour and day of week to assess alignment with adolescent online activity.
    \item \textbf{Platform-side enforcement outcomes}, based on long-term tracking of PHTV availability.  By the writing of this paper in late Jan. 2026, for PHTVs discovered in Jun. 2025, we were able to measure their survival rate across the six-month time span. Then, for PHTVs discovered in Oct. and Dec. 2025, we measured their survival rate over the three-month and one-month time spans accordingly. This availability tracking covered both platforms (Kwai and Douyin) and crucially, we compare the persistence of obfuscated versus unobfuscated PHTVs to quantify the effectiveness of evasion against platform detection.
\end{itemize}

\medskip
\noindent\textbf{{\insight}\label{insight:evasion} PHTV creators systematically disguise harmful intent using low-effort, high-effectiveness obfuscation tactics that exploit gaps in moderation.} 

Through manually annotating 1,755 PHTVs, we find that creators systematically manipulate textual and visual signals to evade automated detectors. We observed two dominant tactics:
(1) \textbf{Semantic Camouflage}: Harmful intent is masked through ironic humor, euphemisms, or “positive” framing. For example, self-harm is described as “dimple art,” and dangerous stunts are labeled as “Legal Channel” content.
(2) \textbf{Noise Injection}: Harmful signals are diluted with irrelevant slogans (e.g., “All professions are valuable”), excessive emojis, or visual effects (e.g., face filters obscuring cigarettes), reducing the salience of risk cues for classifiers.

These tactics are not marginal: 11.05\% of PHTVs employ semantic camouflage, and 23.36\% use noise injection (for examples and more details see Appendix~\ref{appendix:evasion_tactics}). Crucially, they allow harmful content to masquerade as benign, exploiting the gap between surface-level signals and contextual intent.Furthermore, these tactics require minimal effort yet significantly increase content survivability—a finding corroborated by persistence analysis (see Insight~\ref{insight:survival}).

\medskip
\noindent
\textbf{\insight PHTV uploads align with peak adolescent online activity, maximizing initial exposure during high-traffic windows.} 

\begin{figure}
\centering
\includegraphics[width=\linewidth]{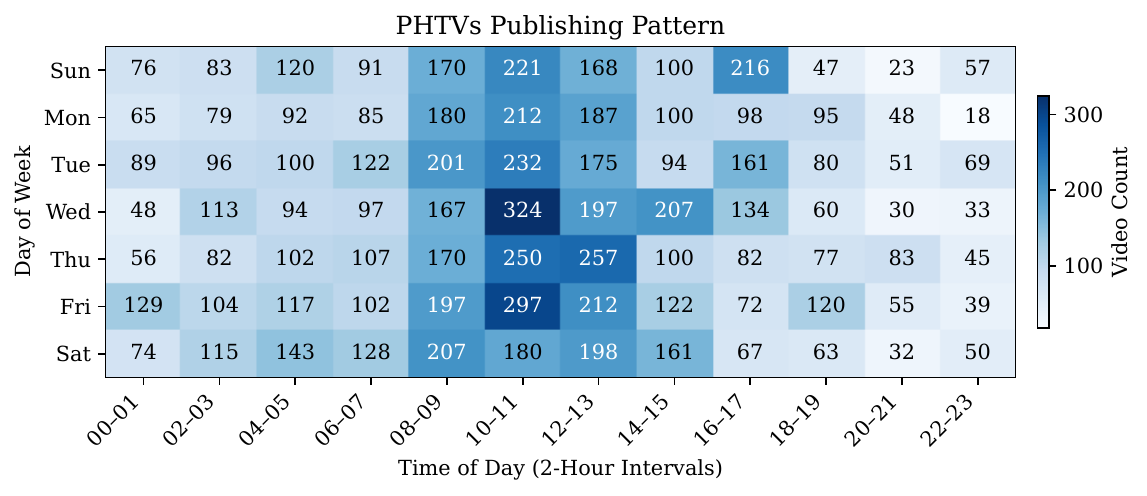}
\caption{Hourly upload pattern of PHTVs (2-hour bins). Peak activity at 10–13 aligns with midday adolescent online presence.}
\label{fig:publish_heatmap}
\end{figure}

This is distilled from the upload timestamps of 10,000 PHTVs collected between Jul. and Dec. 2025. As shown in Figure~\ref{fig:publish_heatmap}, uploads peak sharply during \textbf{10:00–13:00 on weekdays}—aligning precisely with midday breaks in Chinese schools. Weekly patterns further confirm this synchronization: upload volume rises significantly on \textbf{Fridays and Saturdays}, coinciding with extended adolescent free time. While this pattern may partly reflect general creator habits (uploading when they themselves are free), the synchronization with adolescent breaks suggests an emergent or intentional optimization that ensures maximum initial visibility. This temporal alignment acts as an amplification multiplier alongside content-level obfuscation.

\medskip
\noindent
\textbf{\insight\label{insight:survival} Platform enforcement is both delayed and selectively ineffective—particularly against obfuscated content—enabling PHTVs to persist long enough to cultivate harm.} 

This is distilled from longitudinal tracking of PHTV availability: on Kwai, \textbf{95.90\%} of December 2025 PHTVs remained accessible after one month, and \textbf{96.40\%} of October 2025 PHTVs survived three months; on Douyin, \textbf{83.28\%} and \textbf{77.81\%} persisted over the same periods (Table~\ref{tab:phtv_survival_rates}).  
It is also demonstrated by the stark contrast between obfuscated and unobfuscated content: videos lacking evasion tactics (explicit footage of adolescents smoking, drinking, or self-harming, paired with blunt textual descriptions like “Want to drop out,”) are often removed within a month, whereas those using semantic camouflage or noise injection are significantly more likely to survive the critical first month.  
Besides, even after six months, \textbf{38.33\%} of the earliest PHTVs (June 2025 cohort on Kwai) remain publicly available—revealing a systemic latency in long-term content hygiene. Together, these findings confirm an asymmetric arms race: evasion is cheap and adaptive; enforcement is slow and incomplete.

\begin{table}
\centering
\caption{PHTV survival rates on Kwai and Douyin (re-checked in January 2026). 
}
\label{tab:phtv_survival_rates}
\begin{tabular}{l c c c}
\toprule
\textbf{Platform} & \textbf{Collected} & \textbf{Survival} & \textbf{Survival Rate} \\
\midrule
Kwai  & Jun. 2025   & $\geq$6 months & 38.33\% \\
Kwai  & Oct. 2025 & $\geq$3 months & 96.40\% \\
Kwai  & Dec. 2025 & $\geq$1 month & 95.90\% \\
\midrule
Douyin & Oct. 2025 & $\geq$3 months & 77.81\% \\
Douyin & Dec. 2025 & $\geq$1 month & 83.28\% \\
\bottomrule
\end{tabular}
\end{table}

\subsection{Factors Influencing PHTV Exposure in Recommendation Feeds}
\label{subsec:exposure_factors}

Understanding what determines adolescent exposure to PHTVs is essential for effective intervention. If exposure were primarily driven by immutable user attributes—such as age or gender—mitigation would require universal safeguards. However, if it is shaped by modifiable factors—such as platform choice, regulatory context, or transient behaviors—then targeted, scalable solutions become possible. Clarifying this hierarchy of influence directly informs policy design, platform accountability, and digital literacy efforts aimed at empowering adolescents to navigate algorithmic environments safely.

To identify the key drivers of PHTV exposure, we conduct four controlled measurement campaigns:
\begin{itemize}
    \item \textbf{Regulatory impact analysis}: We track PHTV exposure rates for two simulation accounts before and after China's September 2025 crackdown on harmful teen content.
    \item \textbf{Cross-platform comparison}: We deploy identical simulation accounts on Kwai and Douyin (same settings) and measure PHTV prevalence across over 113,000 videos to isolate platform-specific algorithmic effects.
    \item \textbf{Behavioral feedback tracing}: We monitor daily PHTV exposure for a simulation account before and after passive browsing of PHTV comments, capturing short-term algorithmic amplification dynamics.
    \item \textbf{Static attribute variation}: We evaluate exposure by simulation accounts that differ only in declared age (13/15/17), gender, or geographic IP region, while holding behavioral patterns constant.
\end{itemize}

\medskip
\noindent
\textbf{\insight Macro-level regulatory intervention drastically and sustainably reduces PHTV exposure across diverse user profiles.}

This is distilled from longitudinal tracking of two fixed simulated accounts—one cold-start (B) and one mimicking healthy adolescent usage (E)—across July (pre-regulation), October, and December 2025. As shown in Table~\ref{tab:regulation_impact}, PHTV rates plummeted from 14.64\% and 12.92\% in July to 2.90\% and 3.14\% in October—a reduction of over 75\%—and remained stable through December (3.13\% and 3.90\%).  
It is also demonstrated by the consistency of this drop across both account types, confirming that China’s September 2025 nationwide crackdown~\cite{cctv2025} rapidly reshaped the recommendation ecosystem at scale.

\begin{table}
\centering
\caption{PHTV exposure before and after the September 2025 regulatory crackdown.}
\label{tab:regulation_impact}
\begin{tabular}{lccc}
\toprule
\textbf{Account} & \textbf{Month} & \textbf{Total Videos} & \textbf{PHTV Rate} \\
\midrule
B(Cold) & Jul. 2025     & 4,747   & 14.64\% \\
                  & Oct. 2025  & 5,608   & 2.90\%  \\
                  & Dec. 2025 & 1,404    & 3.13\%  \\
\midrule
E(Behavior) & Jul. 2025     & 4,080   & 12.92\% \\
                      & Oct. 2025  & 5,063   & 3.14\%  \\
                      & Dec. 2025 & 1,308   & 3.90\%  \\
\bottomrule
\end{tabular}
\end{table}

\medskip
\noindent
\textbf{\insight Platform choice establishes a persistent baseline disparity in PHTV exposure, independent of user identity or behavior.}  

This is distilled from a controlled comparison using identical synthetic accounts (same age, cold-start, no prior follows) on Kwai and Douyin. As measured over 60,756 videos on Kwai and 52,993 on Douyin, PHTV prevalence was \textbf{3.70\% on Kwai versus 2.65\% on Douyin}—a 1.4× higher exposure rate on Kwai.  
It is also demonstrated by the consistency of this gap across age groups and behavioral variants, indicating that platform-specific content curation policies—not user attributes—drive fundamental differences in risk exposure.

\medskip
\noindent
\textbf{\insight Even minimal, passive user engagement can trigger rapid and severe algorithmic amplification of PHTV exposure.}  

\begin{figure}
\centering
\includegraphics[width=0.65\linewidth]{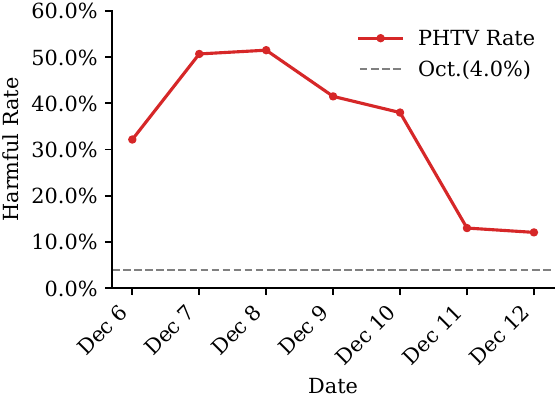}
\caption{Daily PHTV exposure rate for account F after passive browsing of PHTV comments.}
\label{fig:phtv_feedback_loop}
\end{figure}

This is distilled from a 7-day trace of a behaviorally-seeded account F (baseline PHTV rate: 4.01\% in October 2025). On December 6, the account passively browsed comment sections of over 50 representative PHTVs—without liking, sharing, or commenting. As shown in Figure~\ref{fig:phtv_feedback_loop}, this brief interaction caused the daily PHTV rate to surge from 32.16\% on Dec 6 to a peak of \textbf{50.68\% on Dec 7}, remaining elevated ($>$41\%) through Dec 10 before gradually reverting to 12–13\% by Dec 11–12.  
It is also demonstrated by the speed of amplification—within 24 hours—highlighting the extreme sensitivity of recommendation algorithms to transient signals.

\medskip
\noindent
\textbf{\insight Static user attributes—such as declared age, gender, or geographic location—exert negligible influence on PHTV exposure.}  

\begin{figure*}
\centering
\begin{minipage}[b]{0.48\linewidth}
    \centering
    \includegraphics[width=\linewidth]{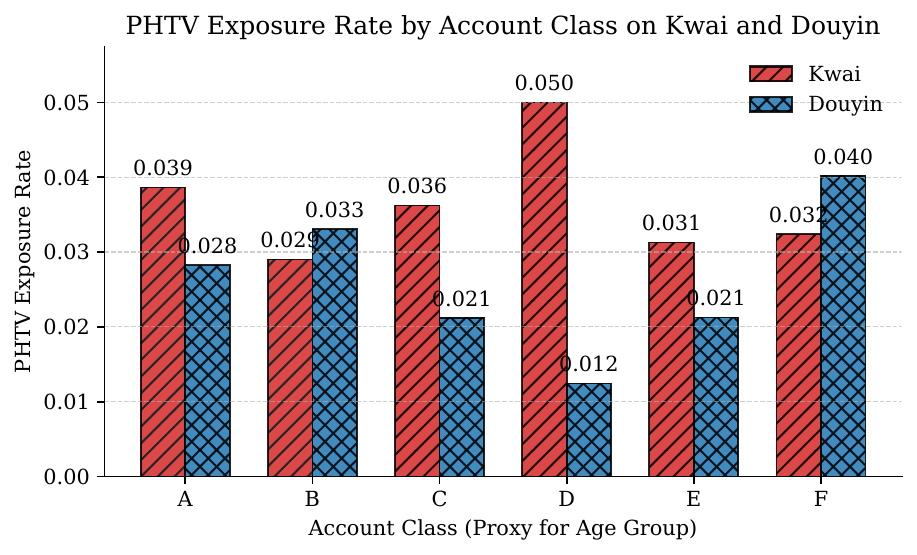}
    \subcaption{PHTV exposure rate by age (13, 15, 17 years) and seeding strategy (cold-start A–C vs. behaviorally-seeded D–F).}
    \label{fig:phtv_rate_by_age_class_a}
\end{minipage}
\hfill
\begin{minipage}[b]{0.48\linewidth}
    \centering
    \includegraphics[width=\linewidth]{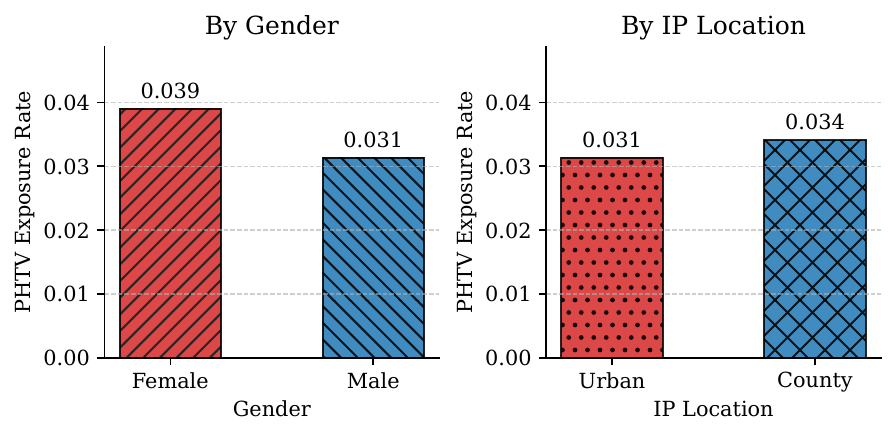}
    \subcaption{PHTV exposure rate by user gender (female/male) and inferred IP location (urban/county).}
    \label{fig:phtv_rate_by_age_class_b}
\end{minipage}
\caption{
    Comparative analysis of PHTV exposure across static identity.
}
\label{fig:phtv_rate_demographic_comparison}
\end{figure*}

This is distilled from testing six synthetic account classes (A–F) that varied only in age (13/15/17 years), gender, or IP region while holding behavioral patterns constant. As shown in Figure~\ref{fig:phtv_rate_by_age_class_a}, PHTV rates show no consistent gradient across age groups or behavioral variants. Similarly, as shown in Figure~\ref{fig:phtv_rate_by_age_class_b}, gender and urban/rural IP configurations yield differences of less than 1 percentage point.  
It is also demonstrated by the stark contrast with dynamic behavioral signals: while passive browsing causes $>$40 pp spikes, changing age or gender alters exposure by almost nothing—confirming that risk is shaped by actions and systems, not identity.

\section{Teenager User Study} \label{sec:teen_user_study}

To ground our measurement framework in authentic adolescent experiences, we conducted an offline questionnaire survey of Chinese middle and high school students. This section outlines the design of the study and the key insights we find. Detailed statistical analyzes are provided in Appendix~\ref{appendix:user_study_detailed}.

\subject{Goals}
The study aimed to: (1) characterize baseline usage patterns of short-video platforms among adolescents aged 12-17; (2) assess adoption rates and perceived utility of ``Youth Mode''; (3) quantify self-reported exposure to inappropriate or harmful content; and (4) identify common interaction behaviors (e.g., liking, sharing) to ground the design of ecologically valid synthetic accounts as a necessary component of our \hunter (Section~\ref{subsec:phtv_hunter}).

\subject{Subject Recruitment}
We conducted an in-class, paper-based survey with trained assistants providing verbal clarification. Two researchers screened responses independently, retaining 683 valid submissions(75.7\% effective response rate) after removing incomplete or low-quality submissions(e.g., missing demographics),  stratified by age: 188 aged 12–13, 224 aged 14–15, and 271 aged 16–17.

\subject{Questionnaire Design.}
Our 13-item questionnaire assesses four core domains: (1) age and platform preference; (2) temporal engagement; (3) interaction behavior; and (4) content safety.The instrument was co-developed with middle school teachers and grounded in China’s minor-protection guidelines to ensure age-appropriateness. Before full deployment, we piloted the survey with 50 students to validate clarity and emotional safety, refining several questions based on their feedback. We use age-appropriate language like "content that made you feel uncomfortable", "dangerous stunts" to ensure adolescent comprehension and minimize distress. We later mapped self-reported exposures to a refined taxonomy (Appendix~\ref{app:survey_mapping} and Appendix~\ref{app:harmful_taxonomy}). The full questionnaire and consent form are provided in Appendix~\ref{app:questionnaire_full} and Appendix~\ref{app:consent_form}.

\subject{Key Insights.}\label{subsec:user_study_key_findings}
Our analysis yields three key insights, each statistically validated and detailed in Appendix~\ref{appendix:user_study_detailed}.

\textbf{\insight Short-video platforms are integral to adolescent life, featuring heavy daily use(especially during holidays) and near-universal engagement through likes, follows, and peer sharing.\label{insight:user_habits}
}

As detailed in Appendix~\ref{app:platform_usage}, Kwai and Douyin dominate the landscape, with over 75\% of respondents naming one as their primary platform. Usage intensity is substantial: even on school days, 27.1\% of 12--13-year-olds spend at least 30 minutes watching videos, while among older teens, heavy usage ($>$2 hours/day) rises to 8.5–10.3\% on school days and surges to 33.5–41.3\% during holidays (Appendix~\ref{app:daily_usage}). Crucially, this consumption is not passive. Interaction is near-universal: liking (54.8–70.1\%), following accounts (38.3–51.3\%), and sharing with peers (45.7–65.6\%) are all prevalent behaviors that increase with age (Appendix~\ref{app:usage_habits}). The proportion of users reporting \textit{no interaction at all} drops from 26.1\% in early adolescence to just 12.9–15.9\% in later years. This confirms that adolescent short-video use is both frequent and socially participatory—providing strong empirical justification for simulating realistic interaction histories in our behavioral-seed accounts.

\textbf{\insight Youth Mode works perfectly—but teens don‘t want it, not due to ignorance, but because teens explicitly reject it as “boring” or “unnecessary.” Safety-by-design fails when users opt out by choice.}

While Youth Mode on both Douyin and Kwai demonstrates near-perfect technical efficacy in blocking PHTV—our measurements show \textbf{0 harmful videos out of 5,863 collected on Douyin and 0 out of 2,688 on Kwai under activated Youth Mode} (Appendix~\ref{appendix:youth_mode})—its real-world impact is severely limited by low adoption. As shown in Appendix~\ref{app:youth_mode}, only 41.5\% of 12--13-year-olds use Youth Mode, declining to 30.6\% among 16--17-year-olds. Among non-users, the primary reasons are not lack of awareness but subjective rejection: 47.3\% to 66.5\% state they “don’t think it’s necessary,” and 31.4\% to 45.4\% find the content “dull” or the restrictions “excessive.” Parental enforcement also drops sharply with age, from 48.4\% to 25.5\%. This reveals a critical disconnect: the protection mechanism works as designed, but adolescents largely opt out of it. Consequently, despite its proven effectiveness, Youth Mode fails to shield the majority of minors who consume content through standard, unfiltered recommendation feeds.

\textbf{\insight Most teens encounter PHTVs, yet most report minimal subjective discomfort; this gap suggests a normalization of harmful content as routine background noise in the feed.}

Specifically, the vast majority of adolescents have been exposed to potentially harmful content on short-video platforms. As shown in Appendix~\ref{app:harmful_taxonomy}, exposure to at least one category of PHTV is widespread: for example, 50.9\% of 14--15-year-olds report encountering life-threatening risk-taking behaviors (C5), 49.1\% see promotion of unhealthy lifestyles (C8), and over 30\% across age groups report exposure to child sexual exploitation imagery (C2), self-harm content (C4), or predatory recruitment (C7). When considering any of the nine harm categories, the cumulative exposure rate exceeds two-thirds of respondents in all age groups.

Despite this high exposure, adolescents exhibit a pronounced disconnect between actual exposure and reported discomfort—i.e., whether they feel the content is inappropriate or makes them uncomfortable. This discrepancy widens among 14-15 year-olds: while over half report seeing specific harmful content, 64.7\% of them indicate they “rarely” or “never” feel uncomfortable when encountering such material (Appendix~\ref{app:awareness_gap}), resulting in a perceived-impact gap of 57.1\%. This discrepancy suggests not an absence of exposure, but rather a normalization of harmful content as routine background noise, reflecting potential habituation or distinct user strategies to actively filter out distressing stimuli.

Furthermore, when asked to rate the personal impact of such content from “no effect at all” (1) to “strong effect” (5), adolescents split into two distinct perceptual profiles (Appendix~\ref{app:impact_clustering}): 66.9\% report consistently low influence on their emotions (1.87), values (1.61), and imitation tendency (1.16), while 33.1\% perceive strong effects across all three dimensions (emotion: 3.93, values: 4.13, imitation: 2.74). This heterogeneity implies that while most teens may downplay the emotional salience of PHTV, a substantial minority experience notable psychological engagement—potentially increasing their susceptibility to behavioral contagion.

Together, these insights provide an empirical basis for our simulation strategy and reveal the realistic status of short-video use among contemporary adolescents in China and the potential threats they face.

\section{Discussion}
\label{sec:discussion}


\subject{Implications for Stakeholders.}

\subsubject{For Platforms:} Our work calls for a shift from reactive takedowns to proactive, context-aware prevention. This includes developing classifiers that can decode semantic camouflage, implementing stricter scrutiny on content uploaded during peak youth activity hours, and redesigning Youth Mode to be more engaging and less paternalistic.

\subsubject{For Policymakers:} China's Sep. 2025 crackdown reduced PHTV prevalence by 75\%, proving the policy's ecosystem-level impact. Yet a persistent harm baseline remains. Future regulations should incentivize platforms not only to build safety features but also to actively drive adoption through improved UX and digital literacy education.

\subsubject{For Researchers \& Educators:} The bimodal distribution of perceived impact—where a substantial minority of teens report strong emotional and behavioral effects—suggests a need for targeted digital literacy programs. These programs should focus on helping teens recognize subtle forms of manipulation, understand the algorithmic forces shaping their feeds, and develop critical resilience against harmful content.

\subject{Limitations and Future Work.} 

\subsubject{Limtations} \textbf{First}, our simulated accounts, though grounded in real survey data, cannot fully replicate the nuanced social graphs and longitudinal histories of real teenagers. This simplified digital footprint may render them more susceptible to platform bot detection mechanisms, potentially biasing the content distribution they trigger. Additionally, our simulation utilizes a finite set of behavioral seeds and controlled session intervals. Real users exhibit significantly higher account diversity, erratic session variability, and sensitivity to initial seed interactions, which could result in more volatile recommendation trajectories than those captured in our measurement. \textbf{Second}, our analysis is limited to two Chinese platforms; the generalizability to other platforms like TikTok or Instagram requires further investigation.\textbf{Third}, our comment analysis, while large-scale, is observational and cannot establish causal links between content exposure and real-world behavior. \textbf{Fourth}, strict API rate limits and data access restrictions on both platforms prevented a complete per-platform breakdown for every analysis. For certain large-scale metrics, we aggregated data across Douyin and Kwai to ensure statistical robustness, prioritizing sample size over platform isolation. \textbf{Finally}, our adolescent survey relied on self-reported data, which may be subject to recall bias. Additionally, placing age-related items at the beginning of the questionnaire may have introduced social-desirability bias, potentially influencing how students reported sensitive experiences regarding harmful content exposure.

\subsubject{Future work} \textbf{First}, we propose the development of a \textit{transparent and decentralized moderation framework}, where guardians (parents) and trusted third parties can co-define personalized safety rules alongside platform and government guidelines. \textbf{Second}, longitudinal studies tracking the same cohort of real teenagers over a long period, while adhering to ethical standards, can provide a deeper understanding of the causal relationship between exposure to harmful video content and developmental outcomes.
\textbf{Third}, expanding our agentic framework to other modalities(eg., live streams) is crucial, as these may be where the most severe predatory interactions occur.


\section{Conclusion}
\label{sec:conclusion}

In this work, we developed \framework, the first large-scale, behaviorally grounded measurement framework for systematically discovering and characterizing PHTVs on China’s dominant short-video platforms. By integrating offline adolescent surveys with online synthetic account simulation, our approach bridges the gap between real-world user behavior and algorithmic exposure, enabling ecologically valid observation of harmful content actually recommended to teens.



Our findings highlight the urgent need for a paradigm shift in youth safety: from reactive takedowns to proactive, context-aware moderation; from one-size-fits-all restrictions to adaptive, engaging safeguards; and from platform-centric solutions to collaborative governance involving regulators, guardians, and researchers. The \framework framework provides both a methodological blueprint and empirical foundation for such efforts, paving the way toward genuinely safer digital spaces for the next generation.


\cleardoublepage
\appendix

\section*{Ethical Considerations}
\label{appendix:ethical_considerations}

\subject{Stakeholders.}
Our research involves multiple stakeholders, including (1) adolescent participants in our offline survey, (2) their legal guardians, (3) the two participating schools (a middle school and a high school), (4) platform users on Kwai and Douyin (including content creators and viewers), (5) the platforms themselves (Kuaishou and Douyin), (6) policymakers, (7) educators, and (8) the broader public concerned with youth digital safety. (9) The research team is also a stakeholder, responsible for ethical data handling and scientific integrity.

\subject{Impacts.}
We considered the following ethical principles from \textit{The Menlo Report}: \textbf{Beneficence}, \textbf{Respect for Persons}, \textbf{Justice}, and \textbf{Respect for Law and Public Interest}.

\noindent\textbf{Harms} include:  
(1) Questions in user studies may make participating teenagers uncomfortable;
(2) potential re-identification or psychological distress if raw video/comment data were leaked;  
(3) unintended amplification of harmful content during account mimicking (e.g., through likes or shares);  
(4) misinterpretation of benign content as harmful due to classification errors;
(5) although our synthetic accounts mimic real teens, they do not represent actual users, raising concerns about representativeness if findings are overgeneralized; and
(6) psychological risks to research team members and annotators who review harmful content during the labeling process.

\subject{Mitigations.}
To address these risks, we implemented multiple safeguards:
\begin{itemize}
    \item \textbf{Questionnaire Design:} The questionnaires used for user surveys were carefully designed to minimize any potential discomfort (see Appendix~\ref{app:questionnaire_full}). The questionnaire clearly stated that participation in this study was entirely voluntary, anonymous, and carried an extremely low risk. No personally identifiable information (such as name, student ID, or phone number) was collected. All data is stored on an encrypted local server, accessible only to research team members, and will be securely deleted after two years.
    
    \item \textbf{Institutional Oversight:} The offline user study received approval from both participating schools, and all participants (and their legal guardians) signed dual-party informed consent forms (see Appendix~\ref{app:consent_form}). The form explicitly stated the low-risk nature of participation and guaranteed anonymity. Our institution does not currently maintain a dedicated IRB-style process for this category of internet measurement and school-based survey work. In the absence of such a mechanism, we followed a documented ethics process grounded in AoIR guidance, school approval, national minor-protection requirements, and ongoing internal risk review.
    
    \item \textbf{Data Minimization \& Anonymization:} We collect only publicly available content from recommendation feeds. All video and comment data are stripped of metadata that could link to specific users and stored on an encrypted, access-controlled server. Raw videos are retained only for validation and will be permanently deleted after publication. We will not release raw video or audio files; any released artifact will be restricted to de-identified, non-invertible, research-oriented representations and documentation, accompanied by content warnings and use restrictions.
    
    \item \textbf{Passive Crawling Protocol:} Our accounts, once set up, perform strictly passive consumption---no likes, comments, shares, or searches are executed---eliminating any risk of algorithmic amplification or engagement with harmful content. This design choice accepts some loss of ecological realism in order to avoid engagement actions that could amplify harmful content.
    
    \item \textbf{Careful Data Collection:} Our data collection on Kwai and Douyin is designed to comply with platform policies and our measurement methodology follows best ethic practices with risks minimized. Particularly, data collection simulated legitimate user behavior using official APIs and rate-limiting delays were enforced to avoid server overload. Besides, content analysis was strictly limited to aggregating statistics and locating predictive insights. All data handling, including the storage of videos and comments, follows protocols for secure and confidential research data management.
    
    \item \textbf{Human-in-the-Loop Validation:} All model-based classifications (binary and multi-class) undergo manual spot-checking by trained annotators. Harmful content examples presented in the paper are fully anonymized (e.g., blurred faces, redacted text). We consider that even blurred screenshots may create reverse-image-search or re-identification risk; therefore, we have removed such images from this version and replaced them with anonymized textual descriptions.
    
    \item \textbf{Responsible Disclosure:} Findings related to platform-specific vulnerabilities (e.g., evasion tactics, timing strategies) are reported at an aggregate level without exposing individual creators or enabling the replication of harmful content. We focused on aggregate disclosure and platform reporting pathways rather than exposing individual creators in the paper. This choice was intended to reduce the risk of harassment, misidentification, and unnecessary secondary harm while still surfacing systemic safety problems.
    
    \item \textbf{Researcher Wellbeing:} Reviewing harmful content poses psychological risks to annotators and researchers. Annotators were informed in advance about the nature of the material, could stop at any time without penalty, worked under capped session lengths (maximum two hours) with mandatory breaks, held debriefs for difficult cases, and had access to institutional counseling support. We established a referral path to university counseling resources for any team member experiencing distress.
    
    \item \textbf{Handling Potentially Illegal Content:} Our taxonomy is intended to study potentially harmful content, not to collect or retain material that is illegal under applicable law. If annotators encountered content suspected of crossing that boundary, they stopped labeling immediately, isolated the case under restricted access, and used the platform's official reporting channel. Such content was not retained as part of the sharable research dataset.
\end{itemize}

\subject{Key Ethical Trade-offs.}
Our study involved several ethical trade-offs that we explicitly acknowledge:

\noindent\textbf{Passive measurement vs. realism.} We accepted some loss of ecological realism in order to avoid engagement actions that could amplify harmful content.

\noindent\textbf{Reproducibility vs. data minimization.} We retained only what was necessary for validation and reproducibility, under encrypted and access-controlled storage, while limiting future sharing to de-identified artifacts.

\noindent\textbf{Systemic reporting vs. individual reporting.} We focused on aggregate disclosure and platform reporting pathways rather than exposing individual creators in the paper. This choice was intended to reduce the risk of harassment, misidentification, and unnecessary secondary harm while still surfacing systemic safety problems.

\subject{Dataset Size and Composition.}
Two datasets served different purposes in our study. The annotated ground-truth set was expanded until the annotation process stopped yielding meaningfully new harmful subtypes. The larger passive measurement corpus was needed to estimate exposure rates and observe lower-frequency patterns such as evasion behaviors. We applied data minimization principles throughout, retaining only content necessary for validation and aggregate analysis.

\cleardoublepage

\section*{Open Science}
The code for this study has been released in an anonymous repository, which you can view here\footnote{\url{https://anonymous.4open.science/r/PHTV_Scout-91B3/}} . This repository contains the complete code for PHTV Scout, as well as the ablation experiment code used to evaluate the effectiveness of the binary classifier, but it does not contain any of the raw dataset content.

To protect user privacy and comply with platform terms of service, we do not release the full raw dataset. However, a sampled and rigorously anonymized subset of the PHTV dataset, including video multimodal features (ASR/OCR transcripts, frames),  categorized labels, and strictly anonymized user study data, will be made available upon request and strict identity validation. Access is contingent upon: (1) institutional affiliation verification, (2) approval by our ethics review board, and (3) signing a data use agreement that explicitly restricts usage to non-commercial academic research and prohibits re-identification attempts or redistribution. 

A final version of the code repository will be provided after the paper is published.
\cleardoublepage
\bibliographystyle{plainurl}
\bibliography{ref}

\appendix
\section{Detailed User Behavior Study Analysis} \label{appendix:user_study_detailed}

This appendix presents a comprehensive analysis of the offline user behavior study conducted among adolescents aged 12--17, which directly informed the design of our synthetic account simulation framework. The survey was administered in person during school hours at one public middle school and one public high school in China. All participants provided written informed consent prior to participation, and data were collected in an anonymous manner without recording any personally identifiable information. Data collection was supervised by trained research team members to ensure accuracy, privacy protection, and ethical compliance.

After removing incomplete responses, inconsistent answering patterns, and participants over the age of 18, a total of 683 valid responses were retained ($n_{12-13} = 188$, $n_{14-15} = 224$, $n_{16-17} = 271$). This appendix details platform usage, temporal engagement patterns, youth mode adoption, exposure to harmful content using an updated nine-category taxonomy, behavioral responses, and cognitive dissonance in harm perception.

\subsection{Platform Usage Patterns} \label{app:platform_usage}

As shown in Table~\ref{tab:platform_usage}, Kwai is the dominant platform across all age groups, with usage rates ranging from 50.6\% to 65.2\%. Douyin usage increases with age, rising from 21.3\% among younger teens to 36.2\% among older adolescents, suggesting a shift toward more curated and trend-driven content as users mature. Xiaohongshu and Bilibili maintain niche roles, with stable but low adoption rates.

\begin{table}
\centering
\caption{Platform preference by age group.}
\label{tab:platform_usage}
\begin{tabular}{@{}lccccc@{}}
\toprule
Age & Douyin & Kwai & Rednote & Bilibili & Other \\
\midrule
12--13  & 21.3\% & 54.3\% & 12.8\% & 6.9\% & 4.8\% \\
14--15  & 18.8\% & 65.2\% & 6.7\% & 6.7\% & 2.6\% \\
16--17  & 36.2\% & 50.6\% & 5.2\% & 6.6\% & 1.5\% \\
\bottomrule
\end{tabular}
\end{table}

\begin{figure}
\centering
\includegraphics[width=\linewidth]{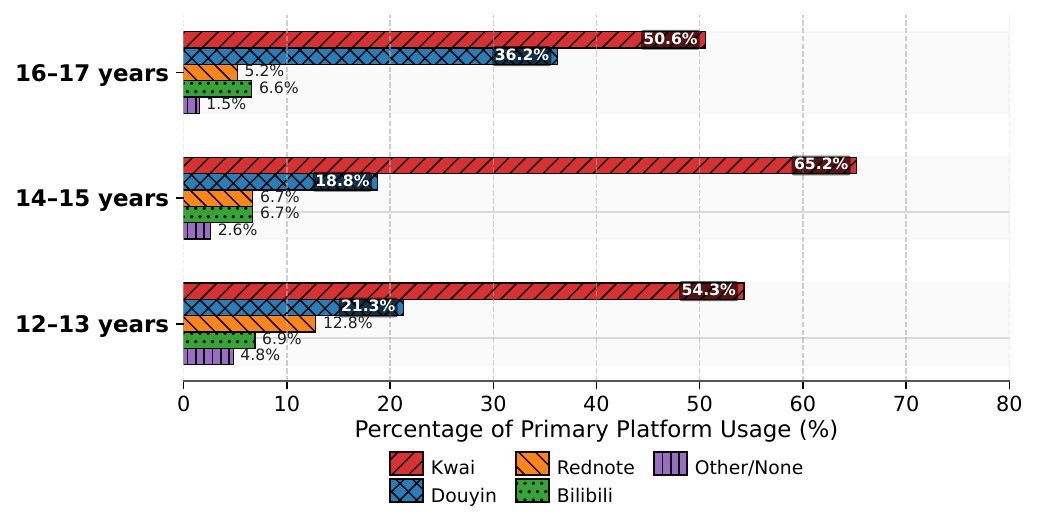}
\caption{Distribution of primary short-video platform preference across age groups.}
\label{fig:platform_chart}
\end{figure}

\subsection{Daily Usage Duration} \label{app:daily_usage}

Usage duration exhibits two key trends: (1) a significant increase during holidays compared to school days, and (2) higher engagement among mid-to-late adolescents. On school days, 72.9\% of 12--13-year-olds use less than 30 minutes or not at all, whereas 14--15-year-olds show the highest weekday engagement. In contrast, holiday usage intensifies across all groups, with heavy usage ($>2$\,hrs/day) increasing from 13.8\% to 41.3\% between the youngest and oldest cohorts.

\begin{table}
\centering
\caption{Heavy usage ($>2$ hrs/day) by age group.}
\label{tab:heavy_usage}
\begin{tabular}{@{}lccc@{}}  
\toprule
Age & School Day & Holiday & Growth Factor \\
\midrule
12--13 years & 4.3\% & 13.8\% & $3.2\times$ \\
14--15 years & 10.3\% & 33.5\% & $3.3\times$ \\
16--17 years & 8.5\% & 41.3\% & $4.9\times$ \\
\bottomrule
\end{tabular}
\end{table}

\begin{figure}
\includegraphics[width=\linewidth]{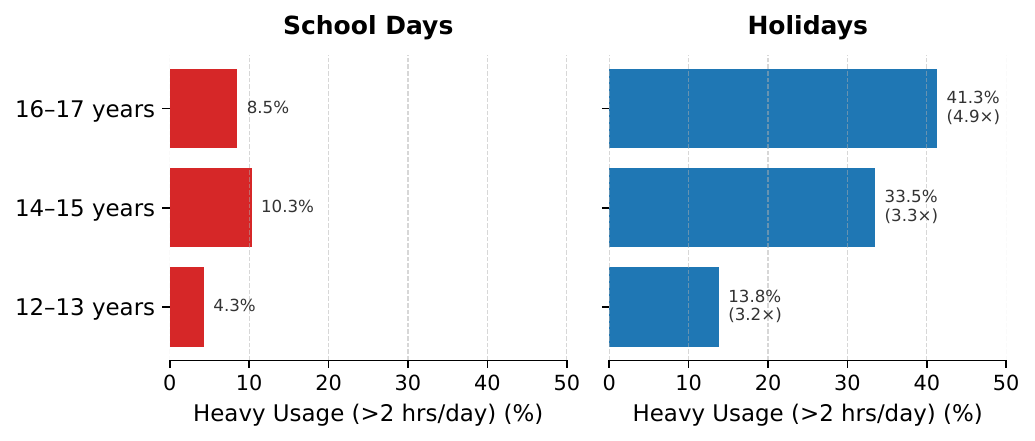}
\caption{Comparison of daily usage duration on school days and holidays across age groups.}
\label{fig:duration_chart}
\end{figure}

\subsection{Youth Mode Adoption and Reasons for Non-Adoption} \label{app:youth_mode}

Youth mode activation declines with age: 41.5\% of 12--13-year-olds report using it, decreasing to 32.1\% and 30.6\% in subsequent groups. Among those who do not use youth mode, the most frequently cited reason is ``don't think it's necessary'' (47.3\%--66.5\%), followed by ``content is dull/uninteresting'' and ``too many feature restrictions.'' Parental supervision is primarily limited to general reminders (45.2\%--59.4\%), with strict limitations declining from 48.4\% to 25.5\% with age.

\begin{table*}
\centering
\caption{Youth mode adoption and key reasons for non-adoption.}
\label{tab:youth_mode}
\begin{tabular}{lccccc}
\toprule
& \multicolumn{2}{c}{Response} & \multicolumn{3}{c}{Top Reasons (among non-users)} \\
\cmidrule(r){2-3} \cmidrule(l){4-6}
Age Group & Yes & No & ``None'' & ``Not Necessary'' & ``Too Restrictive / Dull'' \\
\midrule
12--13 years & 41.5\% & 58.5\% & 70.9\% & 47.3\% & 33.6\%--36.4\% \\
14--15 years & 32.1\% & 67.9\% & 47.4\% & 58.6\% & 44.7\%--45.4\% \\
16--17 years & 30.6\% & 69.4\% & 44.1\% & 66.5\% & 31.4\%--40.4\% \\
\bottomrule
\end{tabular}
\end{table*}

These findings indicate that protection mechanisms are increasingly rejected as adolescents gain autonomy, not due to lack of awareness, but due to perceived irrelevance and restrictive design.

\subsection{Exposure to Harmful Content: A Nine-Category Taxonomy} \label{app:harmful_taxonomy}

We reclassified reported harmful content into a developmentally grounded taxonomy aligned with observed adolescent risks. Each category reflects specific behavioral patterns documented in the survey.

Table~\ref{tab:harmful_exposure} shows the percentage of respondents who reported encountering each category of harmful content.

\begin{table*}
\centering
\caption{Exposure rates to harmful content categories by age group.}
\label{tab:harmful_exposure}
\begin{tabular}{lccc}
\toprule
Harmful Category & 12--13($n=188$) & 14--15($n=224$) & 16--17($n=271$) \\
\midrule
C1: Aberrant Social Media Challenges       & 28.7\% & 29.0\% & 32.8\% \\
C2: Child Sexual Exploitation Imagery      & 19.1\% & 30.4\% & 31.7\% \\
C3: Glorification of Youth Violence        & 22.3\% & 26.3\% & 20.7\% \\
C4: Self-Injury \& Extreme Modifications & 24.5\% & 27.7\% & 17.7\% \\
C5: Life Threatening Risk Taking Behaviors & 39.4\% & 50.9\% & 46.1\% \\
C6: Maladaptive Behavioral Influence        & 34.6\% & 37.9\% & 21.4\% \\
C7: Predatory Recruitment of Minors        & 22.3\% & 31.7\% & 26.9\% \\
C8: Promotion of Unhealthy Lifestyles      & 36.7\% & 49.1\% & 39.5\% \\
C9: Other Harmful                             & 9.0\%  & 8.9\%  & 17.7\% \\
\bottomrule
\end{tabular}
\end{table*}

\begin{figure}
\centering
\includegraphics[width=\linewidth]{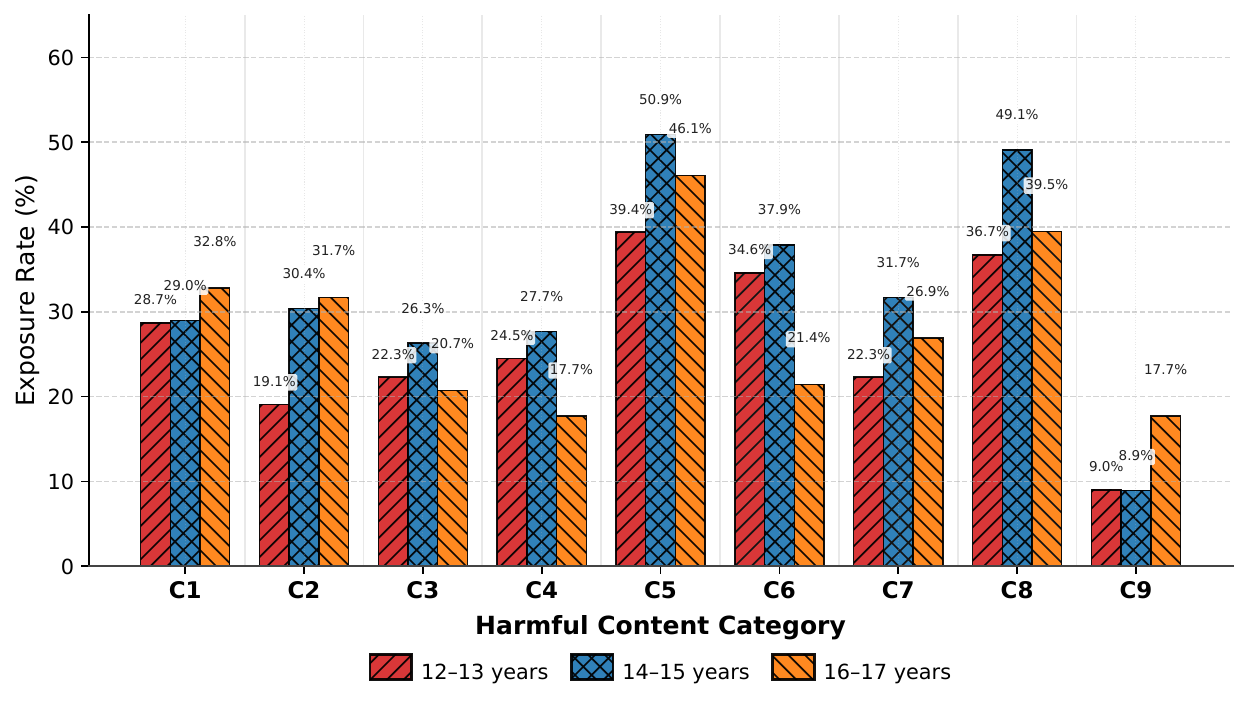}
\caption{Exposure rates to nine categories of harmful content across age groups.}
\label{fig:harm_categories}
\end{figure}

Key observations:
\begin{itemize}
    \item C5 (Life-Threatening Risk-Taking Behaviors) is the most prevalent threat, peaking at 50.9\% among 14--15-year-olds.
    \item C2 (Child Sexual Exploitation Imagery) and C8 (Unhealthy Lifestyles) show strong upward trends with age, indicating increased exposure to adult-themed content.
    \item C4 (Intentional Self-Injury) is most common among younger teens, potentially linked to early mental health crises.
    \item C1 (Aberrant Challenges) rises steadily with age, reaching 32.8\% among 16--17-year-olds.
    \item C9 ("Other") jumps to 17.7\% in the oldest group, suggesting emerging or unclassified digital harms.
\end{itemize}

\subsection{Subjective Awareness vs.\ Actual Exposure} \label{app:awareness_gap}

Despite high actual exposure, subjective awareness remains low. As shown in Table~\ref{tab:awareness_gap}, only a small minority claim never to have seen harmful content, yet a large proportion insist they ``rarely'' or ``never'' feel uncomfortable with all content.

\begin{table*}
\centering
\caption{Discrepancy between actual PHTV exposure and reported discomfort.}
\label{tab:awareness_gap}
\begin{tabular}{lccc}
\toprule
& 12--13 years & 14--15 years & 16--17 years \\
\midrule
Reported ``Rarely'' or ``Never'' feeling uncomfortable & 77.1\% & 64.7\% & 59.4\% \\
Actually never exposed to any PHTV category           & 13.3\% & 7.6\%  & 4.1\% \\
\textbf{Perceived-Impact Gap}                         & \textbf{21.3\%} & \textbf{57.1\%} & \textbf{55.3\%} \\
\bottomrule
\end{tabular}
\end{table*}

The growing gap---peaking at 57.1\% among 14--15-year-olds---indicates progressive desensitization. Many adolescents fail to recognize harmful content as such, treating it as normative entertainment.

\subsection{Behavioral Responses to Harmful Content} \label{app:responses}

When encountering harmful videos, the dominant response is passive avoidance. As shown in Table~\ref{tab:responses}, over 70\% choose to swipe away, while reporting rates remain below 50\% across all groups.

\begin{table}
\centering
\caption{Behavioral responses to harmful videos.}
\label{tab:responses}
\begin{tabular}{lccc}
\toprule
Response Type & 12--13 & 14--15 & 16--17 \\
\midrule
Swipe away           & 78.7\% & 72.8\% & 70.8\% \\
Report               & 41.5\% & 43.8\% & 37.3\% \\
Screenshot \& share  & 5.3\%  & 8.0\%  & 7.4\% \\
Want to know more    & 0.5\%  & 4.5\%  & 7.7\% \\
Feel scared/anxious  & 7.4\%  & 6.2\%  & 3.7\% \\
Don't care           & 6.4\%  & 11.2\% & 16.2\% \\
\bottomrule
\end{tabular}
\end{table}

Notably, emotional distress decreases with age, while curiosity and indifference rise---particularly the desire to ``know more'' (7.7\% in 16--17 group)---suggesting normalization and potential internalization of harmful narratives.

\subsection{Heterogeneity in Perceived Impact of Harmful Content} \label{app:impact_clustering}

To understand how adolescents subjectively evaluate the influence of harmful content, we analyzed responses to Question 12, which asked participants to rate the impact of encountered harmful videos on three dimensions: (1) emotional state, (2) value judgment, and (3) behavioral imitation tendency, using a 5-point Likert scale (1 = no effect, 5 = strong effect).

Applying K-means clustering to the 683 valid minor responses, we identified two distinct subgroups based on silhouette analysis (optimal $K=2$, silhouette score = 0.493). As summarized in Table~\ref{tab:impact_clusters}, the majority cluster (Cluster 0, 66.9\%) reports consistently low perceived impact across all three dimensions (mean scores: 1.87 for emotion, 1.61 for values, 1.16 for imitation), suggesting psychological resilience or desensitization. In contrast, Cluster 1 (33.1\%) exhibits markedly higher sensitivity, with mean scores of 3.93 (emotion), 4.13 (values), and 2.74 (imitation)—indicating that over one-third of adolescents perceive harmful content as substantially affecting their affective and normative frameworks.

\begin{table*}
\centering
\caption{Clustering results of perceived impact (K=2).}
\label{tab:impact_clusters}
\begin{tabular}{lccccc}
\toprule
Cluster & Size (n) & Size (\%) & Emotional State & Value Judgment & Imitation Tendency \\
\midrule
0 & 457 & 66.9\% & 1.87 & 1.61 & 1.16 \\
1 & 226 & 33.1\% & 3.93 & 4.13 & 2.74 \\
\bottomrule
\end{tabular}
\end{table*}

\begin{figure}
\centering
\includegraphics[width=0.85\linewidth]{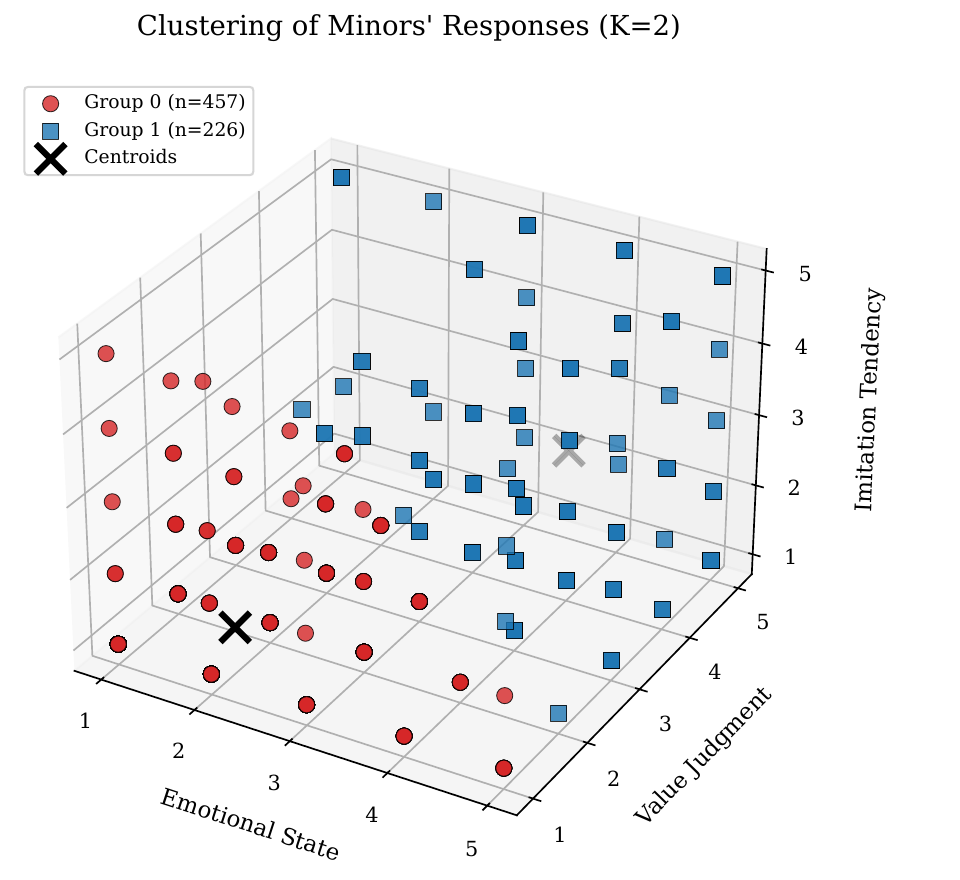}
\caption{3D visualization of adolescent clusters based on perceived impact of harmful content. Cluster 1 (square markers) shows high sensitivity across all dimensions, while Cluster 0 (circle markers) exhibits low impact perception. Black crosses denote cluster centroids.}
\label{fig:impact_clusters_3d}
\end{figure}

\begin{figure}
\centering
\includegraphics[width=0.95\linewidth]{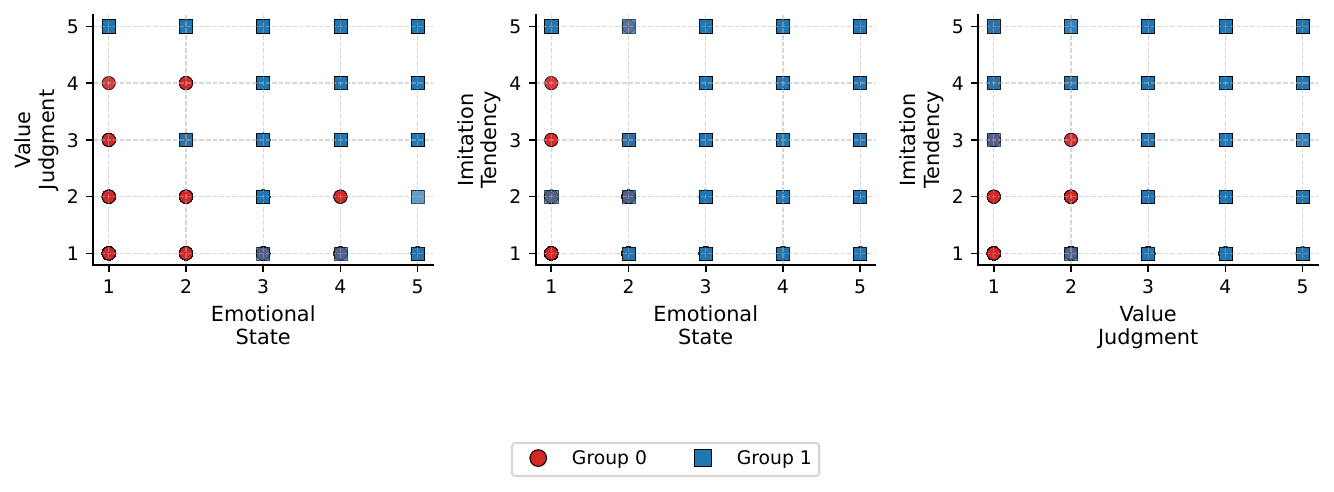}
\caption{2D projections of the same clustering result, showing pairwise relationships between the three impact dimensions. Each subplot reveals clear separation between the two groups, with consistent color and marker encoding (Group 0: red circles; Group 1: blue squares).}
\label{fig:impact_clusters_2d}
\end{figure}

This bimodal distribution highlights a notable heterogeneity in adolescents’ self-reported responses to harmful content: the majority (Cluster 0) report low levels of impact across emotional, value-based, and behavioral dimensions, while a substantial minority (Cluster 1) report consistently higher levels of perceived influence. The difference is most pronounced in value judgment (4.13 vs. 1.61) and emotional state (3.93 vs. 1.87), with a moderate gap also observed in imitation tendency (2.74 vs. 1.16). Both the 3D and 2D visualizations confirm the robustness of this dichotomy, underscoring that a significant fraction of teenagers are highly susceptible to the affective and normative influence of PHTV—highlighting the urgent need for targeted interventions.

\subsection{Usage Habits} \label{app:usage_habits}

Beyond platform choice and exposure to harmful content, understanding adolescents' interactive behaviors on short-video platforms is critical for designing ecologically valid simulation accounts. As shown in Table~\ref{tab:usage_habits}, engagement patterns vary significantly across age groups, with notable increases in social interaction and content sharing among older teens.

The most common interaction behavior across all age groups is \textit{liking} videos, ranging from 54.8\% of 12--13-year-olds to 69.0\% of 16--17-year-olds. Following accounts is also prevalent, increasing from 38.3\% to 46.5\% with age, indicating growing investment in personalized content curation.

Sharing videos with friends emerges as a dominant social practice, rising sharply from 45.7\% (12--13 years) to 63.1\% (16--17 years), reflecting the increasing role of short videos in peer communication and identity formation during late adolescence. Commenting remains relatively stable at around 35\%--48\%, suggesting that while many consume and share content, active textual participation is less common.

Notably, the proportion of users who report \textit{no interaction} decreases with age: from 26.1\% among 12--13-year-olds to 15.9\% among 16--17-year-olds, indicating greater behavioral engagement over time. A non-trivial minority---up to 37.2\%---report watching most videos completely, suggesting high attention retention despite concerns about fragmented attention spans.

These findings support our decision to simulate light but realistic interaction histories (e.g., liking 10--20 videos, following 30--50 accounts) for behavioral-seed accounts, aligning with observed user habits without simulating extreme or atypical engagement.

\begin{table*}
\centering
\caption{Interactive usage habits by age group.}
\label{tab:usage_habits}
\begin{tabular}{lccc}
\toprule
Behavior & 12--13 years ($n=188$) & 14--15 years ($n=224$) & 16--17 years ($n=271$) \\
\midrule
Like                        & 54.8\% & 70.1\% & 69.0\% \\
Follow account              & 38.3\% & 51.3\% & 46.5\% \\
Comment                     & 37.8\% & 47.8\% & 34.7\% \\
Share with friends          & 45.7\% & 65.6\% & 63.1\% \\
Watch most videos completely& 37.2\% & 44.2\% & 38.0\% \\
No interaction              & 26.1\% & 12.9\% & 15.9\% \\
\bottomrule
\end{tabular}
\end{table*}

\begin{figure}
\centering
\includegraphics[width=\linewidth]{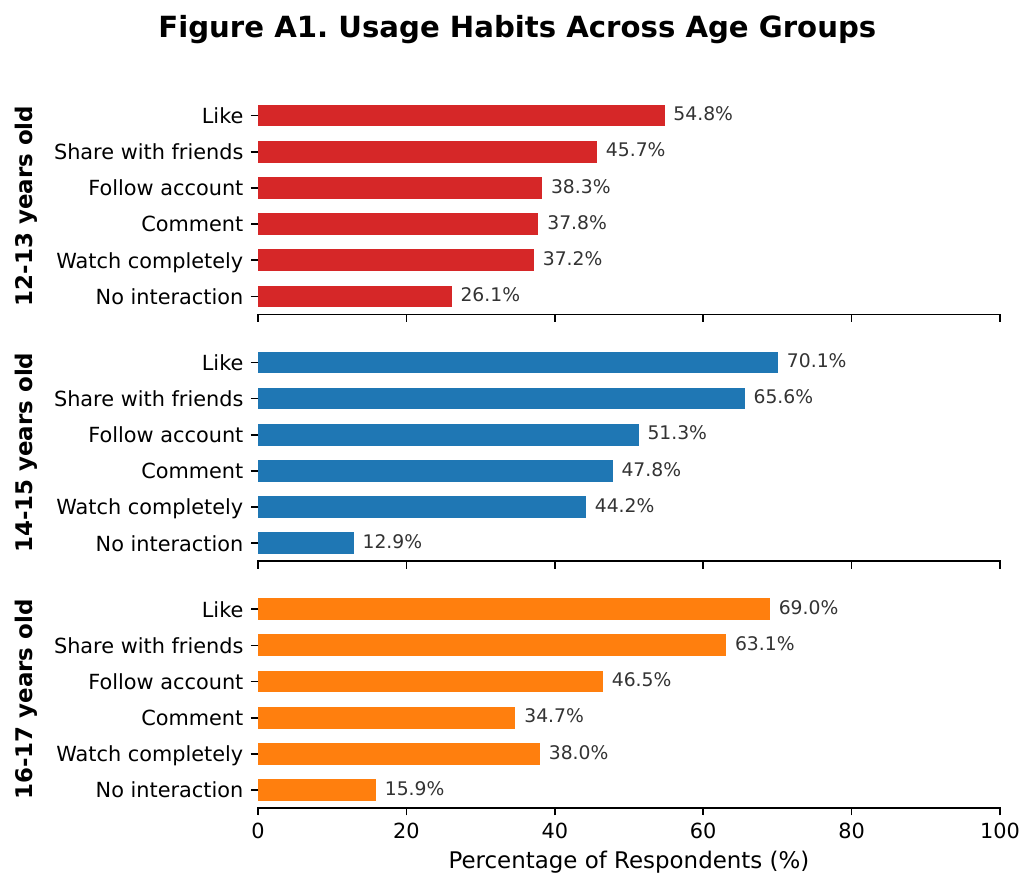}
\caption{Distribution of interactive behaviors across age groups.}
\label{fig:usage_habits}
\end{figure}

\subsection{Implications for Account Design and Ethical Execution} \label{app:implications}

All data collection and simulation tasks were executed on a dedicated \textbf{Linux server}, using isolated Android emulators to host individual accounts. Device fingerprints were standardized, and all traffic routed through a fixed Beijing-based IP address to control for geographic and device-level biases.

Account configurations were derived directly from survey results:
\begin{itemize}
    \item \textbf{Cold-start accounts}: Simulate new users with no interaction history.
    \item \textbf{Behavioral-seed accounts}: Perform calibrated interactions (liking $\sim$15 videos, following $\sim$40 accounts) based on observed habits.
    \item \textbf{Stratification}: By age group and gender (randomized).
    \item \textbf{Passive consumption only}: No sensitive searches; all content from recommendation feed.
\end{itemize}

Ethical safeguards included:
\begin{itemize}
    \item Informed consent from all participants and guardians;
    \item Data anonymization and secure storage on encrypted drives;
    \item Compliance with platform terms of service (no automation violating ToS);
    \item Minimal request frequency to avoid server overload.
\end{itemize}

This rigorous foundation ensures both ecological validity and methodological integrity in our investigation of algorithmic exposure to harmful content.

\subsection{Full Questionnaire Instrument} \label{app:questionnaire_full}

The following is the complete English translation of the paper-based questionnaire administered to adolescent participants. All responses were collected anonymously.

\bigskip

\noindent\textbf{Brief Survey on Adolescent Short-Video Platform Usage}

\smallskip

\noindent Note: This survey is for academic research only. All responses are anonymous and will be kept strictly confidential. Please answer based on your actual experience. Thank you for your support!

\begin{enumerate}
    \item What is your age? \\
    \underline{\hspace{4cm}}

    \item Which short-video platform do you use most often? (Single choice) \\
    $\square$ Douyin \quad $\square$ Kwai \quad $\square$ Bilibili \quad $\square$ Rednote \quad $\square$ Other: \underline{\hspace{2cm}} \quad $\square$ None

    \item On school days, approximately how much time do you spend watching short videos? (Single choice) \\
    $\square$ Almost never \quad $\square$ Less than 30 minutes \quad $\square$ 30 minutes – 1 hour \quad $\square$ 1–2 hours \quad $\square$ More than 2 hours

    \item During holidays, approximately how much time do you spend watching short videos? (Single choice) \\
    $\square$ Almost never \quad $\square$ Less than 30 minutes \quad $\square$ 30 minutes – 1 hour \quad $\square$ 1–2 hours \quad $\square$ More than 2 hours

    \item Do you perform any of the following actions on short-video platforms? (Multiple choices) \\
    $\square$ Like videos \quad $\square$ Follow accounts \quad $\square$ Comment \quad $\square$ Share with friends \quad $\square$ Watch most videos to completion (without swiping away) \quad $\square$ Never interact

    \item Have you currently enabled ``Youth Mode''? \\
    $\square$ Yes \quad $\square$ No

    \item If you have not enabled Youth Mode, what are the main reasons? (Multiple choices) \\
    $\square$ Content is too dull/uninteresting \quad $\square$ Too many feature restrictions (e.g., cannot comment or go live) \quad $\square$ Don’t know where to enable it \quad $\square$ Don’t think it’s necessary \quad $\square$ Parents didn’t require it \quad $\square$ Other: \underline{\hspace{2cm}}

    \item Have you noticed that “after watching a certain video, similar content keeps being recommended”? \\
    $\square$ Never noticed \quad $\square$ Occasionally \quad $\square$ Often \quad $\square$ Always

    \item While browsing short videos, have you ever encountered content that made you feel uncomfortable or inappropriate? \\
    $\square$ Never \quad $\square$ Rarely \quad $\square$ Sometimes \quad $\square$ Often \quad $\square$ Almost every time

    \item Which of the following types of content have you seen? (Multiple choices) \\
    $\square$ Soft pornography (e.g., revealing clothing, suggestive gestures) \\
    $\square$ Topics related to body modification, self-harm, depression, or suicide \\
    $\square$ Dangerous stunts or reckless driving \\
    $\square$ Cyberbullying, physical violence, or bloody/gory scenes \\
    $\square$ Smoking, drinking, or betel nut chewing \\
    $\square$ Scams, fake job offers targeting minors, or suspicious “earn money online” schemes \\
    $\square$ Content promoting food waste, property destruction, or dropping out of school \\
    $\square$ Extreme ugliness challenges or vulgar/low-quality pranks \\
    $\square$ Other: \underline{\hspace{2cm}}

    \item When you encounter such content, how do you usually respond? (Multiple choices) \\
    $\square$ Swipe away immediately \quad $\square$ Report the video \quad $\square$ Take a screenshot and share with friends \quad $\square$ Feel scared or anxious \quad $\square$ Don’t care \quad $\square$ Want to know more (keep watching)

    \item How much do you think this type of content affects you? (1 = No effect at all, 5 = Strong effect) \\
    Emotional state: $\square$ 1 $\square$ 2 $\square$ 3 $\square$ 4 $\square$ 5 \\
    Value judgments: $\square$ 1 $\square$ 2 $\square$ 3 $\square$ 4 $\square$ 5 \\
    Tendency to imitate behaviors: $\square$ 1 $\square$ 2 $\square$ 3 $\square$ 4 $\square$ 5

    \item Do your parents restrict your use of short-video apps? \\
    $\square$ Strictly restricted (e.g., time limits, supervision) \quad $\square$ General reminders \quad $\square$ Rarely intervene \quad $\square$ No restrictions at all
\end{enumerate}

\subsection{Mapping from Survey Language to PHTV Taxonomy}
\label{app:survey_mapping}

To ensure the questionnaire was age-appropriate, we used descriptive, relatable language rather than academic or clinical terms. Table~\ref{tab:survey_taxonomy_mapping} details the mapping between the options presented in the survey (Question 10) and the refined PHTV taxonomy used for our analysis.

\begin{table*}
\centering
\caption{Mapping between questionnaire descriptions and PHTV taxonomy categories.}
\label{tab:survey_taxonomy_mapping}
\begin{tabular}{p{0.55\textwidth} p{0.35\textwidth}}
\toprule
\textbf{Questionnaire Description (Question 10)} & \textbf{PHTV Taxonomy Category} \\
\midrule
``Soft pornography (e.g., revealing clothing, suggestive gestures)'' & \textbf{C2:} Child Sexual Exploitation Imagery \\
\addlinespace
``Cyberbullying, physical violence, or bloody/gory scenes'' & \textbf{C3:} Glorification of Youth Violence \\
\addlinespace
``Topics related to body modification, self-harm, depression, or suicide'' & \textbf{C4:} Intentional Self-Injury \& Extreme Body Modifications \\
\addlinespace
``Dangerous stunts or reckless driving'' & \textbf{C5:} Life-Threatening Risk-Taking Behaviors \\
\addlinespace
``Content promoting food waste, property destruction, or dropping out of school'' & \textbf{C6:} Maladaptive Behavioral Influence \\
\addlinespace
``Scams, fake job offers targeting minors, or suspicious ``earn money online'' schemes'' & \textbf{C7:} Predatory Recruitment of Minors \\
\addlinespace
``Smoking, drinking, or betel nut chewing'' & \textbf{C8:} Promotion of Unhealthy Lifestyles \\
\addlinespace
``Extreme ugliness challenges or vulgar/low-quality pranks'' & \textbf{C1:} Aberrant Social Media Challenges \\
\addlinespace
``Other: \underline{\hspace{2cm}}'' & \textbf{C9:} Other Digital Harm to Minors \\
\bottomrule
\end{tabular}
\end{table*}

\subsection{Informed Consent Form} \label{app:consent_form}

The following is the English translation of the dual-signature informed consent form provided to participants and their legal guardians prior to survey administration.

\bigskip

\noindent\textbf{Informed Consent Form for Adolescent Short-Video Usage Survey}

\medskip

\noindent Dear Student, \\
Hello!

We are conducting an academic research study on adolescents’ short-video platform usage behaviors, aiming to understand your daily habits and experiences with these platforms. The findings will help improve content recommendation systems and inform better youth protection policies.

You are invited to participate in this brief questionnaire survey. Please read the following information carefully.

\begin{enumerate}
    \item \textbf{Study Content} \\
    The questionnaire will ask about your short-video usage, including preferred platforms, time spent, interaction behaviors, Youth Mode adoption, and exposure to uncomfortable or inappropriate content. It contains 13 questions and takes approximately 5 minutes to complete.

    \item \textbf{Voluntary Participation} \\
    Your participation is entirely voluntary. You may stop at any time without any negative consequences.

    \item \textbf{Anonymity and Confidentiality} \\
    No personally identifiable information (e.g., name, phone number, ID) will be collected. \\
    All data will be anonymized and used solely for academic analysis. \\
    Data will be kept strictly confidential and will not be disclosed to any third parties.

    \item \textbf{Risks and Benefits} \\
    Risk: Very low. The survey does not involve sensitive or psychologically distressing topics. \\
    Benefit: Your responses will contribute to improving youth protection mechanisms on short-video platforms, with positive societal impact.

    \item \textbf{Data Storage and Retention} \\
    All data will be used for research reports, academic publications, or policy recommendations. Data will be stored in encrypted format and securely deleted two years after the study concludes.

    \item \textbf{Contact Information} \\
    If you have any questions about this study, please contact the research team via the institutional email provided by the ethics review board.
\end{enumerate}

\bigskip

\noindent I have read and understood the above information and voluntarily agree to participate in this survey.

\bigskip

\noindent Participant signature: \underline{\hspace{5cm}} \\
Parent/guardian signature: \underline{\hspace{5cm}} \\
Date: \underline{\hspace{2cm}} Year \underline{\hspace{1.5cm}} Month \underline{\hspace{1.5cm}} Day

\section{Binary Classification for PHTV Detection} \label{appendix:binary_classifier_details}

This appendix provides a comprehensive exposition of the binary classification module, a critical component within the PHTV Hunter framework. It details the model selection rationale, dataset construction methodology, training procedure, performance evaluation, and deployment strategy, offering full technical transparency for the results reported in Section~\ref{subsec:phtv_arbiter}.

\subsection{Model Selection and Architecture Justification}\label{appendix:model_select}
To balance strong multimodal reasoning capabilities with the practical limitations of consumer-grade GPUs—particularly restricted VRAM that allows only one model to run at a time—we conducted an extensive comparative analysis across multiple state-of-the-art vision-language models. We evaluated both open-source models such as \textit{Qwen2.5-VL-7B-Instruct} and \textit{Gemma3-12B}, which can be deployed locally under tight memory budgets, as well as powerful proprietary models including \textit{Qwen-VL-MAX}, \textit{GPT-4-Turbo}, and \textit{Claude-Opus-4.1}. Initial experiments focused on in-context learning (ICL) with varying shot numbers ($k$) aiming to identify approaches that achieve high accuracy without requiring large-scale model parallelism or excessive memory footprint.

The ICL evaluations revealed that while zero-shot performance was promising, it exhibited significant variance and lower recall, particularly for subtle or contextually complex harmful content. This motivated us to pursue fine-tuning on a domain-specific dataset curated from adolescent-facing content. Among all evaluated models, \textbf{Qwen3-VL-8B-Instruct} demonstrated the most favorable trade-off between performance gain from fine-tuning and computational efficiency. Its architecture proved highly effective in integrating visual, textual, and temporal cues from video data.

To achieve parameter-efficient fine-tuning (PEFT) and minimize trainable parameters, we employed \textbf{Low-Rank Adaptation (LoRA)}. The LoRA adapter was configured with a rank $r=8$ and scaling factor $\alpha=16$, which allowed us to update only a small subset of the model's parameters while preserving its pre-trained knowledge, thus drastically reducing memory footprint and training time.

We further compared two versions of the LoRA-finetuned Qwen3-VL-8B-Instruct classifier, differing primarily in prompt language and input configuration. The final version, featuring a Chinese prompt and enhanced multimodal inputs (larger image resolution, integrated ASR/OCR), significantly outperformed its predecessor.


\begin{table*}
\centering
\caption{Performance comparison across all evaluated models. All ICL experiments used the same complex prompt (with full category definitions and descriptions) as the final binary classifier. The final LoRA-finetuned Qwen3-VL-8B-Instruct model serves as the baseline.}
\label{tab:all_model_comparison}
\begin{tabular}{lccccc}
\toprule
Model & Accuracy (\%) & Precision (\%) & Recall (\%) & F1-Score (\%) & FPR (\%) \\
\midrule
\multicolumn{6}{c}{\textit{Multimodal Fusion Models}} \\
\addlinespace[0.5em]
Early Fusion (Text+Video)              & 92.57 & 94.08 & 90.86 & 92.44 & 5.71 \\
Late Fusion (Text+Video)               & 77.43 & 86.92 & 64.57 & 74.10 & 9.71 \\
BERT (Text-only)                       & 78.57 & 76.88 & 81.71 & 79.22 & 24.57 \\
VideoMAE (Video-only)                  & 72.29 & 71.91 & 73.14 & 72.52 & 28.57 \\
\addlinespace[1em]
\multicolumn{6}{c}{\textit{In-Context Learning Experiments}} \\
\addlinespace[0.5em]
Gemma3-12B                             & 53.43 & 87.50 & 8.00  & 14.66 & 1.14 \\
Gemma3-4B                              & 57.14 & 59.54 & 44.57 & 50.98 & 30.29 \\
Qwen2.5-VL-3B-Instruct                 & 46.86 & 41.79 & 16.00 & 23.14 & 22.29 \\
Qwen2.5-VL-7B-Instruct                 & 68.86 & 81.73 & 48.57 & 60.93 & 10.86 \\
Qwen-VL-MAX (128-shot)                 & 79.66 & 82.10 & 76.00 & 78.93 & 16.67 \\
Qwen-VL-MAX + CoT-CN                   & 63.87 & 94.12 & 30.57 & 46.15 & 1.96 \\
Qwen3-VL-Plus + CoT-CN                 & 58.00 & 96.67 & 16.57 & 28.29 & 0.57 \\
Qwen-VL-MAX + CoT-EN                   & 58.60 & 93.55 & 17.16 & 29.00 & 1.15 \\
GPT-4-Turbo + CoT-EN                   & 54.29 & 66.67 & 17.14 & 27.27 & 8.57 \\
Claude-Opus-4-1 + CoT-EN               & 63.87 & 89.83 & 30.81 & 45.89 & 3.45 \\
\addlinespace[1em]
\multicolumn{6}{c}{\textit{Final Deployed Model}} \\
\addlinespace[0.5em]
\textbf{Qwen3-VL-8B-Instruct (LoRA)}   & \textbf{94.29} & \textbf{96.41} & \textbf{92.00} & \textbf{94.15} & \textbf{3.43} \\
\bottomrule
\end{tabular}
\end{table*}

\subsection{Dataset Construction and Annotation Protocol}
\label{appendix:ground_truth}
Our binary classification dataset comprises 3,510 videos, evenly balanced between 1,755 \texttt{Harmful} and 1,755 \texttt{Benign} samples. Importantly, the \texttt{Harmful} category is not composed of generic inappropriate content, but specifically includes videos that either feature adolescents as primary subjects or target adolescents as their intended audience. These videos were collected via recommendation feeds triggered by simulation accounts configured with adolescent-age profiles and behaviorally seeded interactions. While synthetic accounts cannot fully replicate the social graphs and viewing histories of real teenagers, this approach approximates the recommendation pathways through which adolescents may encounter harmful content.

The ground-truth dataset construction followed an iterative, snowball-based seeding process to discover PHTV-relevant keywords and authors, followed by behavioral simulation to trigger personalized recommendations:

\begin{enumerate}
    \item \textbf{Initial Discovery}: We deployed passive simulation accounts to browse recommendation feeds without any interaction. From the recommended videos, annotators identified initial PHTV instances and extracted recurring terms from video titles, ASR transcripts, OCR text, and hashtags.
    
    \item \textbf{Keyword Snowball Expansion}: Each extracted term was used as a search query. For all videos returned by a query, we crawled the authors' full public video histories and identified additional PHTVs. New recurring terms from these videos were added to the keyword pool. This process was iterated until no new keywords emerged across two consecutive rounds (keyword saturation).
    
    \item \textbf{Behavioral Simulation}: Using the saturated keyword set and identified authors, we created dedicated simulation accounts that systematically \textbf{followed}, \textbf{liked}, and \textbf{favorited} content from these authors. This mimicked the behavioral pattern of an engaged teenage user who has shown interest in similar content. These accounts were used exclusively for ground-truth construction and were deleted after annotation.
    
    \item \textbf{Recommendation Feed Harvesting}: After establishing the simulated behavioral profile, we scraped the personalized recommendation feeds across five iterative collection rounds ($\sim$2,000 videos per round, $\sim$10,000 total). The final dataset of 3,510 videos (1,755 harmful, 1,755 benign) was selected from this pool through rigorous manual annotation. Data collection ceased after the final two rounds yielded no new harmful subcategories, confirming taxonomy saturation.
    
    \item \textbf{Manual Annotation}: Three trained graduate student annotators with expertise in child safety and digital media independently labeled all videos using the refined taxonomy (Section~\ref{sss:multi_classification}).
\end{enumerate}

\begin{itemize}
    \item \textbf{Taxonomy Development}: The nine-category taxonomy was grounded in China's \textit{Regulations on the Protection of Minors on the Internet} and refined through iterative review of collected PHTVs. An initial codebook was tested on a 200-video pilot study, and category definitions were updated based on edge cases identified by annotators.
    \item \textbf{Annotation Guidelines}: The refined codebook defined each category with concrete, behaviorally anchored examples (Table~\ref{tab:taxonomy}). Key decision boundaries were established: C1 (socially deviant but non-dangerous) vs.\ C5 (substantial injury/death risk); C4 (deliberate self-harm as the goal) vs.\ C5 (harm as a byproduct); C6 (undermines social functioning) vs.\ C8 (health-compromising behaviors).
    \item \textbf{Annotator Training and Calibration}: Three trained annotators completed a 4-hour training session using 100 pre-labeled exemplar videos. Weekly calibration meetings were held throughout annotation to address emerging edge cases.
    \item \textbf{Inter-Annotator Agreement (IAA)}: Pairwise Cohen's $\kappa$ exceeded 0.90 across all annotator pairs on the full 1,755-video dataset. Discrepancies were resolved through structured discussion with a senior researcher.
    \item \textbf{Data Splitting}: The dataset was partitioned into a train-test split, resulting in 3,160 training samples and 350 test samples.
\end{itemize}


\subsection{Training Procedure and Hyperparameters}
The LoRA-finetuned Qwen3-VL-8B-Instruct model was trained using a custom script leveraging DeepSpeed for efficient distributed training. The key hyperparameters and configurations are summarized below:

\begin{itemize}
    \item \textbf{Base Model}: \texttt{Qwen3-VL-8B-Instruct}
    \item \textbf{Fine-Tuning Method}: Low-Rank Adaptation (LoRA)
    \item \textbf{LoRA Configuration}: Rank $r=8$, Scaling Factor $\alpha=16$
    \item \textbf{Learning Rate}: $1 \times 10^{-4}$
    \item \textbf{Batch Size}: 1 per device, with gradient accumulation over 4 steps (effective batch size = 4)
    \item \textbf{Epochs}: 2
    \item \textbf{Video Sampling}: Frames sampled at 0.5 FPS, with a maximum of 30 frames per video.
    \item \textbf{Input Resolution}: Video pixel count dynamically scaled within a range (min: 784, max: 50176) to optimize for both quality and computational load.
    \item \textbf{Hardware}: Training was performed on a machine equipped with two NVIDIA GeForce RTX 4090 GPUs.
    \item \textbf{Optimization}: AdamW optimizer, cosine learning rate scheduler, with a warmup ratio of 0.03.
\end{itemize}

\subsection{Multi-feature Engineering}\label{appendix:asr_ocr}

\subject{Automatic Speech Recognition (ASR)}
We utilize the \texttt{FireRedASR-AED\cite{xu2025fireredasr}} model to transcribe the audio track of each video into text. The audio is extracted using \texttt{ffmpeg}, converted to mono 16kHz WAV format, and processed with beam search decoding. The maximum audio duration to process is configurable (default: 60 seconds).

\subject{Optical Character Recognition (OCR)} We employ \texttt{PaddleOCR\cite{cui2025paddleocr30technicalreport}} to detect and recognize text within video frames. Frames are sampled at a configurable interval frame rate $t_f$ (default: $t_f = 2$ seconds). To minimize redundancy from static subtitles, we apply sequence similarity matching (\texttt{SequenceMatcher}) to retain only novel text additions across frames.

\subsection{Prompt Engineering}\label{appendix:prompt}
\begin{figure*}
\centering
\fbox{%
  \rule{0pt}{1em}
  \begin{minipage}{0.96\textwidth}
    \centering
    \includegraphics[width=\textwidth]{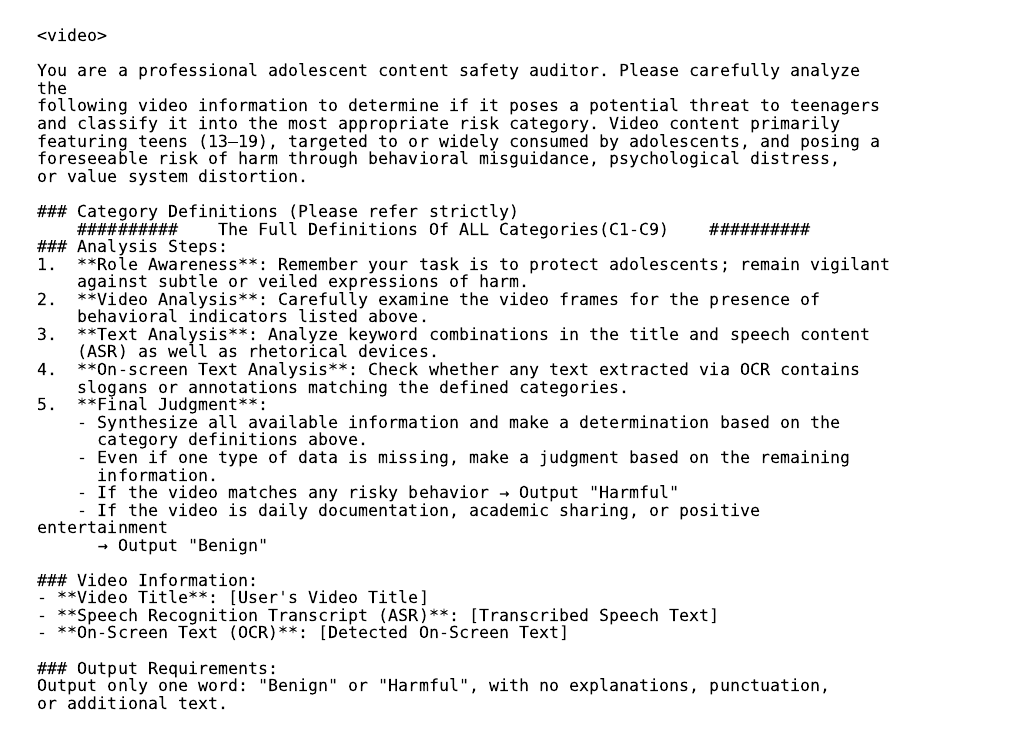}
  \end{minipage}%
  \rule{0pt}{1em}
}
\caption{Template structure of the training prompt for the binary classifier. [Note: This figure illustrates the layout of the prompt, showing where the video title, ASR, OCR, frame sequences, and category definitions are placed relative to the instruction.]}
\label{fig:prompt_template}
\end{figure*}

A structured prompt in Chinese was designed to guide the model. The prompt includes the full definition of the nine harmful categories (as per Table~\ref{tab:taxonomy}) and instructs the model to act as a professional moderator. The complete structure of the training prompt, including the placement of multimodal inputs, is illustrated in Figure~\ref{fig:prompt_template}.

\subsection{Performance Evaluation}\label{appendix:performance_and_eval}
The final LoRA-finetuned \textbf{Qwen3-VL-8B-Instruct} model was selected for deployment based on its superior performance, as summarized in Table~\ref{tab:all_model_comparison}.

To investigate the impact of temporal sampling density and aggregation strategies on classification robustness, we conducted an ablation study across varying frame sampling frequencies and decision fusion rules. Given that Qwen3-VL-8B-Instruct processes video inputs by explicitly specifying the frames-per-second (fps) parameter—rather than accepting arbitrary frame sequences—we designed our experiments around controlled fps configurations. For multi-frame evaluation, we adopted a fixed set of four sampling rates: \textbf{0.3, 0.5, 0.7, and 1.0 fps}, denoted as \textbf{4fps}. At each fps setting, the model independently analyzes the corresponding subset of frames extracted from the video (e.g., one frame every $\sim$3.3 seconds at 0.3 fps). This approach enables systematic comparison of temporal coverage versus performance trade-offs.

To investigate the impact of temporal sampling density and aggregation strategies on classification robustness, we conducted an ablation study across varying frame sampling frequencies and decision fusion rules. Given that Qwen3-VL-8B-Instruct processes video inputs by explicitly specifying the frames-per-second (fps) parameter—rather than accepting arbitrary frame sequences—we designed our experiments around controlled fps configurations. For multi-frame evaluation, we adopted a fixed set of four sampling rates: \textbf{0.3, 0.5, 0.7, and 1.0 fps}, denoted as \textbf{4fps}. At each fps setting, the model independently analyzes the corresponding subset of frames extracted from the video (e.g., one frame every $\sim$3.3 seconds at 0.3 fps). This approach enables systematic comparison of temporal coverage versus performance trade-offs.

Three aggregation strategies were evaluated over the 4fps configuration:
\begin{itemize}
    \item \textbf{Single FPS}: Uses prediction from only one fps value (here, 0.5) to represent lightweight inference.
    \item \textbf{Majority Vote}: Final label determined by majority vote across all four fps predictions.
    \item \textbf{Any-Harmful}: If any single fps prediction yields \texttt{Harmful}, the final decision is \texttt{Harmful}, maximizing recall.
\end{itemize}

We further extended this analysis across different observation durations (10s, 20s, 30s), extracting frames accordingly based on the specified fps. The results are summarized in Table~\ref{tab:ablation_study_reorganized}.

As shown, the \texttt{any harmful} strategy consistently achieves the highest recall (up to 96.57\%) and F1-score (94.41\%), demonstrating its sensitivity in detecting subtle or intermittently presented harmful signals. However, it incurs a higher FPR (8.00\% at 10s) and quadruples inference time due to multiple model calls, making it less suitable for large-scale deployment. In contrast, the \texttt{majority vote} strategy offers marginal improvements in precision but suffers from increased latency with diminishing returns in accuracy.

Among all configurations, \texttt{single fps of 10s} strikes an optimal balance: achieving 93.43\% accuracy, 95.24\% precision, and an average inference time of just \textbf{1.13 seconds}. Its efficiency stems from analyzing only two representative frames within the first 10 seconds of the video, significantly reducing computational load while preserving strong discriminative power. Moreover, this configuration aligns well with typical user engagement patterns on short-video platforms, where initial frames heavily influence recommendation dynamics.

Considering the constraints of real-world deployment—including throughput requirements, GPU memory limitations, and the need for daily batch processing—we ultimately adopt the \texttt{single fps of 10s} configuration as the operational standard for large-scale measurement. While slightly below the full-model baseline in recall, its high precision and sub-second latency enable scalable, sustainable monitoring across thousands of videos per day.

\begin{table*}
\centering
\caption{Ablation study on frame sampling frequency and aggregation strategy. Performance metrics are expressed as percentages. The baseline is the final deployed LoRA-finetuned \textbf{Qwen3-VL-8B-Instruct} model. Multi-frame strategies use the 4fps configuration: \{0.3, 0.5, 0.7, 1.0\} fps.}
\label{tab:ablation_study_reorganized}
\begin{tabular}{lcccccc}
\toprule
Configuration & Accuracy (\%) & Precision (\%) & Recall (\%) & F1-Score (\%) & FPR (\%) & Time (s) \\
\midrule
\multicolumn{7}{c}{\textit{\textbf{Baseline}}} \\
\addlinespace[0.5em]
\textbf{Qwen3-VL-8B-Instruct (LoRA)} & \textbf{94.29} & \textbf{96.41} & \textbf{92.00} & \textbf{94.15} & \textbf{3.43} & \textbf{2.54} \\
\addlinespace[1em]

\multicolumn{7}{c}{\textit{Single-Frame Sampling (fps=0.5)}} \\
\addlinespace[0.5em]
single fps of 10s    & 93.43 & 95.24 & 91.43 & 93.29 & 4.57 & 1.13 \\
single fps of 20s    & 93.71 & 96.36 & 90.86 & 93.53 & 3.43 & 1.61 \\
single fps of 30s    & 93.71 & 95.27 & 92.00 & 93.60 & 4.57 & 1.79 \\
\addlinespace[1em]

\multicolumn{7}{c}{\textit{Multi-Frame Majority Voting (fps=4: \{0.3, 0.5, 0.7, 1.0\})}} \\
\addlinespace[0.5em]
majority voting of 10s  & 93.43 & 95.24 & 91.43 & 93.29 & 4.57 & 5.41 \\
majority voting of 20s  & 93.43 & 95.78 & 90.86 & 93.26 & 4.00 & 7.45 \\
majority voting of 30s  & 93.14 & 95.21 & 90.86 & 92.98 & 4.57 & 8.34 \\
\addlinespace[1em]

\multicolumn{7}{c}{\textit{Any-Harmful Detection (fps=4: \{0.3, 0.5, 0.7, 1.0\})}} \\
\addlinespace[0.5em]
any harmful of 10s     & 94.29 & 92.35 & 96.57 & 94.41 & 8.00 & 5.41 \\
any harmful of 20s     & 94.29 & 93.79 & 94.86 & 94.32 & 6.29 & 7.45 \\
any harmful of 30s     & 94.29 & 93.79 & 94.86 & 94.32 & 6.29 & 8.34 \\
\bottomrule
\end{tabular}
\end{table*}

To evaluate the generalization capability of the deployed model under realistic conditions, we conducted an additional test on a manually labeled dataset of 1,000 wild samples collected from live crawling sessions. This out-of-distribution evaluation aims to reflect the model’s performance in practical deployment scenarios, where content distribution, presentation styles, and noise levels differ from curated training data.

The results demonstrate strong real-world effectiveness: the model achieves an accuracy of 96.80\%, with a precision of 95.60\% and a recall of 97.95\%, yielding an F1 score of 96.76\%. The false positive rate (FPR) is measured at 4.30\%, indicating a low tendency to misclassify benign content as harmful. Notably, the recall increases significantly compared to the internal test set (97.95\% vs. 92.00\%), suggesting that the model is highly sensitive to diverse and evolving manifestations of PHTV content in real-world streams. At the same time, the high precision confirms its ability to maintain specificity despite distributional shifts.

These results demonstrate that the LoRA-finetuned Qwen3-VL-8B-Instruct classifier not only performs well on held-out validation data but also maintains robustness and stability when exposed to real-world, uncurated video streams. This validates its suitability for large-scale, long-term monitoring of PHTV content on short-video platforms.

\section{Multi-class Classifier Details} \label{appendix:multi_classifier_details}

This appendix provides technical details of the fine-grained multi-class classification system described in Section~\ref{sss:multi_classification}, including dataset construction, model configuration, evaluation metrics, and a comparative analysis between LoRA-finetuned models and in-context learning (ICL) on \textit{Qwen-VL-MAX}. The comparison highlights that while ICL achieves competitive accuracy, its high latency renders it impractical for large-scale deployment.

\subsection{Dataset Construction}

The multi-class classifier operates exclusively on videos labeled as \texttt{Harmful} by the binary detection model. A total of 1,755 harmful videos were manually annotated into eight predefined categories (\texttt{C1}--\texttt{C8}), with no instances assigned to \texttt{C9} (Other Harmful), indicating clear category separation during annotation. Category definitions align with those used in the binary dataset composition analysis (see Table~\ref{tab:binary_dataset_composition}) and are grounded in adolescent developmental risks.

Using stratified sampling, the dataset was partitioned into a training set of 1,255 samples and a test set of 530 samples to preserve class distribution. Final labels were determined through consensus among three annotators.

\subsection{Model Configuration}
\label{appendix:multi_model_configuration}

Following the same methodology as the binary classifier, we employ a LoRA-finetuned \textbf{Qwen3-VL-8B-Instruct} model with rank $r=8$ and $\alpha=16$, trained on multimodal inputs: video frames sampled at 0.5\,FPS, ASR transcripts, OCR-extracted text, and video titles. Training was conducted on two NVIDIA GeForce RTX 4090 GPUs using the AdamW optimizer with a learning rate of $2 \times 10^{-4}$ over 10 epochs.

The prompt template extends the binary classification structure---retaining full category definitions and behavioral descriptions---but expands the output space to support prediction across \texttt{C1} to \texttt{C8}. For visualization of the prompt layout, refer to Figure~\ref{fig:prompt_template}.

\subsection{Performance Evaluation of the LoRA-Finetuned Model}

The LoRA-finetuned model demonstrates strong generalization across all eight harmful content categories. Per-class precision, recall, F1-score, and support on the test set ($n = 530$) are reported in Table~\ref{tab:multi_class_performance}.

\begin{table}
\centering
\caption{Per-class performance of the LoRA-finetuned multi-class classifier ($n = 530$).}
\label{tab:multi_class_performance}
\begin{tabular}{@{}lcccc@{}}
\toprule
\textbf{Category} & \textbf{Precision} & \textbf{Recall} & \textbf{F1-Score} & \textbf{Num} \\
\midrule
C1 & 88.89\% & 80.00\% & 84.21\% & 30 \\
C2 & 94.18\% & 98.89\% & 96.48\% & 180 \\
C3 & 71.43\% & 71.43\% & 71.43\% & 21 \\
C4 & 92.86\% & 100.00\% & 96.30\% & 39 \\
C5 & 96.55\% & 94.92\% & 95.73\% & 59 \\
C6 & 91.89\% & 91.89\% & 91.89\% & 37 \\
C7 & 97.98\% & 100.00\% & 98.98\% & 97 \\
C8 & 96.49\% & 82.09\% & 88.71\% & 67 \\
\midrule
\textbf{Accuracy} & \multicolumn{4}{r}{$93.96\%$} \\
\textbf{Wei. Avg.} & 93.97\% & 93.96\% & 93.85\% & — \\
\textbf{Mac. Avg.} & 91.28\% & 89.90\% & 90.47\% & — \\
\bottomrule
\end{tabular}
\end{table}

The model achieves an overall accuracy of \textbf{94\%}, with high F1-scores for well-represented classes such as \texttt{C2}, \texttt{C4}, \texttt{C7}, and \texttt{C5}. Slightly lower performance on \texttt{C1} and \texttt{C8} reflects challenges in distinguishing nuanced behavioral cues, while \texttt{C3}'s limited sample size constrains generalization. Nevertheless, the model is robust enough for integration into the PHTV Analyzer pipeline.

\subsection{Comparative ICL Experiments with Qwen-VL-MAX}

To evaluate zero-shot alternatives without fine-tuning, we conducted \textbf{128-shot} in-context learning (ICL) experiments using the state-of-the-art \textit{Qwen-VL-MAX} model via API access. The same prompt structure, input modalities, and evaluation protocol were applied, refer to Figure~\ref{fig:prompt_template}. One instance was filtered out by the API’s safety mechanism, resulting in a final test set of $n = 529$.

While the ICL approach achieved a respectable accuracy of \textbf{91.68\%}, its average inference time exceeded \textbf{10 seconds per video}, compared to just \textbf{2.54 seconds} for the LoRA-finetuned model---a more than fourfold increase in latency. This delay becomes prohibitive when scaling to thousands of daily queries.

Moreover, despite slightly improved recall for underrepresented classes (e.g., +6.67 pp for \texttt{C1}), ICL exhibited significant drops in precision for others (e.g., $-25.89$ pp for \texttt{C6}), leading to reduced macro-averaged F1-score. These fluctuations suggest instability in reasoning consistency under complex prompts.

To facilitate direct performance comparison, Table~\ref{tab:performance_difference} presents the difference in per-class metrics between the LoRA-finetuned model and the ICL method (\textbf{LoRA $-$ ICL}), expressed in percentage points (pp). Positive values indicate superiority of the LoRA model; negative values indicate relative weakness. Note that support counts differ slightly due to one sample being filtered in the ICL run.

\begin{table}
\centering
\caption{Difference in per-class performance (LoRA $-$ ICL)}
\label{tab:performance_difference}
\begin{tabular}{@{}lcccc@{}}
\toprule
\textbf{Category} & \textbf{Precision} & \textbf{Recall} & \textbf{F1-Score} & \textbf{Num} \\
\midrule
C1 & +12.42 pp & -6.67 pp & +2.96 pp & 30 \\
C2 & -4.68 pp & +2.22 pp & -1.27 pp & 180 \\
C3 & -21.90 pp & +4.76 pp & -6.35 pp & 21 \\
C4 & +6.19 pp & 0.00 pp & +3.44 pp & 39 \\
C5 & +3.22 pp & -1.63 pp & +0.81 pp & 59 \\
C6 & +25.89 pp & +2.70 pp & +16.03 pp & 37 \\
C7 & +0.04 pp & +2.06 pp & +1.04 pp & 97 \\
C8 & +4.18 pp & +10.45 pp & +8.04 pp & 67 \\
\midrule
\textbf{Accuracy} & \multicolumn{4}{r}{+2.28 pp} \\
\textbf{Wei. Avg.} & +1.40 pp & +2.28 pp & +2.16 pp & — \\
\textbf{Mac. Avg.} & +3.17 pp & +1.73 pp & +3.09 pp & — \\
\bottomrule
\end{tabular}
\end{table}

Key observations from the differential analysis include:
\begin{itemize}
    \item \textbf{Precision}: LoRA significantly outperforms ICL in precision for most categories, especially \texttt{C6} (+25.89 pp) and \texttt{C8} (+4.18 pp), while suffering only minor losses in \texttt{C2} (-4.68 pp) and \texttt{C3} (-21.90 pp).
    \item \textbf{Recall}: ICL shows higher recall in some rare classes like \texttt{C1} (+6.67 pp advantage for ICL), but LoRA maintains better balance overall, particularly excelling in \texttt{C8} (+10.45 pp).
    \item \textbf{F1-Score}: The LoRA model achieves superior or comparable F1 across all classes, with notable gains in \texttt{C6} (+16.03 pp) and \texttt{C8} (+8.04 pp), contributing to a +3.09 pp improvement in macro F1.
    \item \textbf{Aggregate Metrics}: Across weighted and macro averages, LoRA consistently leads in both recall and F1, confirming its stronger generalization capability, especially for imbalanced data.
\end{itemize}

\subsection{Conclusion: Deployment Considerations}

Although ICL on \textit{Qwen-VL-MAX} shows promising zero-shot capability, its substantial inference latency, dependency on external services, and inferior macro-level performance make it unsuitable for our large-scale measurement framework. In contrast, the LoRA-finetuned \textbf{Qwen3-VL-8B-Instruct} offers superior efficiency, deterministic behavior, and better overall metric stability---making it the preferred choice for deployment in the PHTV Analyzer system.

Thus, we adopt the LoRA-finetuned model as the default multi-class classifier in our pipeline.

\section{Video Collection}
\label{appendix:video_collection}

We maintain login sessions for simulated teenager accounts to periodically fetch their personalized video feeds from platform recommendation systems. To avoid detection as abnormal accounts or overloading servers, we employ session rotation, randomized request timing, and immediate termination upon CAPTCHA or access denial. Additionally, a dual-delay strategy mimics natural browsing behavior: a 4–8 second delay between consecutive requests simulates typical video viewing duration, while a 10–15 minute pause after every 30 successfully retrieved videos emulates natural usage breaks and reduces server load.

Regarding crawling personalized video feeds,  platform-specific official web endpoints are implemented. On \textit{Kwai}, data is collected using the platform's GraphQL interface, specifically the \texttt{visionNewRecoFeed} query with dynamically updated \texttt{pcursor} parameters for paginated streams. On \textit{Douyin}, content is retrieved via a RESTful endpoint responsible for delivering personalized video feeds, with GET requests incorporating headers and query parameters derived from authenticated mobile app sessions.

\section{Longitudinal PHTV Tracking}
\label{appendix:longtitudinal_tracking}

\textit{PHTV Engagement Tracking.}
To capture user engagement on an PHTV, the system automatically retrieves the comment section of each newly detected harmful video. For each platform, we utilize its official API to fetch up to 500 top-level comments and their associated replies. This ensures high-fidelity collection of textual interactions, including commenter sentiment, peer influence patterns, and geographic origin (e.g., IP location labels on Douyin).

To preserve ecological validity and avoid detection, the crawler employs a dual-delay strategy:
\begin{itemize}
    \item A small delay of 5--10 seconds after each request or page turn, simulating natural reading behavior.
    \item A large delay of several minutes after processing every 20–30 videos, emulating extended breaks in user activity.
\end{itemize}
Furthermore, all data is organized in a structured manner: comments are stored in account-specific directories (\texttt{comments\_data/<account\_name>/}), with each video's discussion saved as an isolated CSV file named by its unique identifier. An anti-duplication mechanism prevents redundant retrieval, ensuring that only fresh content is collected once per day.

\textit{PHTV Persistence Tracking.}
To assess the lifespan of PHTVs, we implement a periodic liveness tracker. The tracker periodically sends lightweight HTTP GET requests to the public URL of each identified harmful video, verifying whether it remains accessible. If the response indicates a successful load (HTTP 200), the system records the video’s continued existence; if the page returns a 404 or redirection (e.g., to a takedown notice), the video is marked as removed.

The entire tracking pipeline operates under strict concurrency control using a thread pool, allowing parallel execution across accounts while adhering to platform rate limits. All operations are timestamped and failure-tolerant, supporting manual resumption in case of interruption.

\section{Sentiment Analysis and Thematic Clustering of PHTV Comments}
\label{appendix:sentiment_thematic}

This appendix details the methodology employed for sentiment analysis and thematic clustering of user comments under videos classified as Potentially Harmful to Teens and Youth (PHTV). These analyses aim to capture both the emotional valence and semantic patterns of public discourse surrounding PHTV content, thereby enabling large-scale and longitudinal observation of its societal impact.

\subsection{Sentiment Analysis}

\textbf{Objective.} To assess the emotional tone of comments associated with PHTV videos, facilitating macro-level analysis of public sentiment toward such content.

\textbf{Model and Approach.} We leveraged \texttt{Qwen3-Max-Instruct}, a state-of-the-art large language model, via in-context learning (ICL). The model was prompted to classify each comment into one of three predefined sentiment categories: \textit{positive}, \textit{negative}, or \textit{neutral}, without fine-tuning.

\textbf{Prompt Design.} The ICL prompt explicitly defined category semantics and enforced strict JSONL output formatting to ensure programmatic parsing. Key instructions included:
\begin{itemize}
    \item Clear definitions for each sentiment class (e.g., \textit{positive} includes praise, encouragement; \textit{neutral} includes factual statements or mentions with~‘@’).
    \item Requirement to output per-comment results with index alignment and brief justifications.
    \item Prohibition of any extraneous text beyond the JSONL lines.
\end{itemize}

\textbf{Validation and Performance.} We evaluated the model on a human-annotated test set of 1,000 comments, randomly selected from 3,111 comments across eight randomly choosed PHTVs (covering all content categories). The results in table~\ref{tab:sentiment_performance} demonstrated high reliability:

\begin{table}
\centering
\caption{Sentiment analysis performance of \texttt{Qwen3-Max-Instruct} (ICL).}
\label{tab:sentiment_performance}
\begin{tabular}{lccc}
\toprule
\textbf{Class} & \textbf{Precision} & \textbf{Recall} & \textbf{F1-Score} \\
\midrule
Positive  & 1.0000 & 0.9836 & 0.9917 \\
Negative  & 0.9830 & 0.9665 & 0.9746 \\
Neutral   & 0.9845 & 0.9937 & 0.9891 \\
\midrule
Macro Avg & 0.9891 & 0.9813 & 0.9852 \\
Weighted Avg & 0.9870 & 0.9870 & 0.9870 \\
\bottomrule
\end{tabular}
\end{table}

The overall accuracy reached 98.70\%, confirming the model’s suitability for large-scale deployment in our pipeline.

\subsection{Thematic Clustering}

\textbf{Initial Attempts.} We first experimented with unsupervised methods (HDBSCAN, K-means) on the same 5,000-comment corpus. However, these approaches yielded clusters of low semantic coherence due to the brevity, noise, and high variability of short-form video comments.

\textbf{Adopted Strategy.} We instead employed a \textit{prompt-engineered, model-assisted semantic clustering} approach using \texttt{Qwen3-Max-Instruct}. For each PHTV category, we processed batches of 200–1,000 comments through a structured prompt that required the model to:
\begin{itemize}
    \item Induce 3–5 interpretable thematic categories.
    \item Provide a name, description, and three representative examples per theme.
    \item Explicitly isolate all comments containing ‘@’ into a dedicated category: \textit{Social Summoning and User Mention (@xxx)}.
    \item Assign every comment to a category via indexed mapping.
\end{itemize}

This human-in-the-loop design enabled high-level semantic abstraction while preserving critical platform-specific behaviors (e.g., user mentions).

\textbf{Emergent Themes.} The analysis identified a set of recurrent thematic categories across all PHTV types. Representative examples originally in Chinese are provided with English translations in parentheses:

\begin{table*}
\centering
\caption{Thematic categories identified in PHTV comment analysis. (TC means Thematic Category)}
\label{tab:thematic_clusters}
\begin{tabularx}{\textwidth}{@{}>{\raggedright\arraybackslash}p{0.28\textwidth}>{\raggedright\arraybackslash}X@{}}
\toprule
\textbf{Theme Name} & \textbf{Description \& Example (Chinese with English translation)} \\
\midrule
TC1: Supportive Feedback & Encouragement or validation of creators. \\
 & \textit{Example: ``So cool!!!''} \\
\midrule
TC2: Critical Feedback & Skepticism about authenticity or ethics. \\
 & \textit{Example: ``Why waste [time/money]????''} \\
\midrule
TC3: Hostile Language & Abusive, threatening, or discriminatory remarks. \\
 & \textit{Example: ``If you keep posting videos like this that spread toxic ideas, I'll f***ing kill you''} \\
\midrule
TC4: Imitative Behavior Expression & Explicit intent or desire to replicate behaviors shown. \\
 & \textit{Example: ``I want to drill a few holes in my face [cool girl]''} \\
\midrule
TC5: Social Summoning and User Mention (@xxx) & Use of `@' to reference other users. \\
 & \textit{Example: ``@xxx''} \\
\midrule
TC6: Resonance-Driven Self-Disclosure & Personal narratives triggered by emotional resonance; often trauma- or experience-based. \\
 & \textit{Example: ``I was also this tough when I was little! But adulthood and reality trained me to never hit back or talk back!''} \\
\midrule
TC7: Predatory Recruitment Targeting Minors & Solicitation for risky activities (e.g., debt, illegal gigs). \\
 & \textit{Example: ``Anyone want to be a punching bag? 288 RMB/hour''} \\
\midrule
TC8: Promotion of Risky Lifestyles & Marketing of substances like e-cigarettes or tobacco. \\
 & \textit{Example: ``Selling cigarettes—like this post if you need some''} \\
\midrule
TC9: Sexual Harassment and Contact Solicitation Involving Minors & Sexualized comments or requests for private contact, often targeting young creators. \\
 & \textit{Example: ``Little sister, do you have WeChat? Follow me back''} \\
\midrule
TC10: Other / Noise & Uninterpretable, off-topic, or nonsensical text. \\
 & \textit{Example: ``Seven Dragons Six Phoenixes''} \\
\bottomrule
\end{tabularx}
\end{table*}

These themes inform our subsequent quantitative and qualitative investigations into how PHTV content shapes—and is shaped by—user interactions, particularly among adolescent audiences.

\textbf{Ethical Safeguards.} All comment data underwent anonymization prior to analysis. No personally identifiable information was retained, and theme labels deliberately avoid linking expressions to individual identities. The inclusion of ``Resonance-Driven Self-Disclosure'' reflects our commitment to recognizing vulnerable user behaviors while ensuring such data is handled with heightened confidentiality.

\section{Evasion Techniques of PHTV}
\label{appendix:evasion_tactics}

In this section, we will detail two different strategies for circumventing platform content moderation.

\subsection{Semantic Camouflage}
\label{subsec:semantic_camouflage}

We identify \textit{semantic camouflage} as a prevalent and sophisticated evasion strategy in which creators deliberately substitute explicit harmful concepts with seemingly innocuous, metaphorical, or culturally coded terms—thereby preserving the underlying meaning for target audiences while evading lexical and semantic detectors. This tactic exploits the gap between surface-level text and contextual intent, allowing harmful content to masquerade as benign, humorous, or even educational.

Through systematic annotation of 1,755 PHTV videos collected from real adolescent recommendation feeds, we uncover 13 distinct types of jargon (or “coded language”) that function as semantic substitutes across multiple PHTV categories. These expressions are not random slang but stable, community-recognized euphemisms that consistently map to specific harmful behaviors. Crucially, such coded terms appear in the textual metadata (titles, captions, on-screen text, or ASR transcripts) of at least 11.05\% of all PHTV videos in our dataset—providing a conservative lower-bound estimate of semantic camouflage prevalence.

Table~\ref{tab:semantic_camouflage_examples} presents the identified jargon lexicon, including the original Chinese term, its literal English translation, the associated PHTV category, and a contextual description of its actual meaning. Notably, many terms repurpose everyday phrases (e.g., “broadcast calisthenics,” “roller skating”) or positive idioms (e “reading ten thousand books...”) to normalize or glamorize dangerous, illegal, or self-destructive acts. This repackaging not only bypasses keyword filters but also leverages cultural familiarity to reduce viewer skepticism—particularly among adolescents who may decode the hidden meaning through peer context.

The widespread use of such semantic substitutes underscores a critical limitation of current moderation systems: models that rely solely on explicit language or surface semantics will systematically miss content that is \textit{intentionally designed to be ambiguous to algorithms yet clear to humans}. Effective detection thus requires contextual understanding, cross-modal alignment (e.g., linking “nine-tailed fox” to images of multiple ear piercings), and continuous lexicon updating—a challenge that demands both technical and sociolinguistic insight.

\begin{table}
\centering
\caption{Examples of semantic camouflage jargon in PHTV.}
\label{tab:semantic_camouflage_examples}
\begin{tabularx}{\linewidth}{@{}>{\raggedright\arraybackslash}Xc>{\raggedright\arraybackslash}X@{}}
\toprule
\textbf{Euphemism} & \textbf{Category} & \textbf{Actual Meaning / Description} \\
\midrule
Good life & C1 & Eating on the ground with strange behavior and actions. \\
Broadcast calisthenics & C2 & Minors performing sexually suggestive dances. \\
Nine-tailed fox & C4 & Getting multiple ear piercings. \\
Dimple & C4 & A form of self-harm involving facial piercings. \\
Paronychia challenge & C4 & Smashing objects onto one’s feet to attract attention. \\
Series connection & C5 & Connecting flammable items (e.g., alcohol lamps) end-to-end and igniting them. \\
Roller skating & C5 & Riding only on the rear wheel of a non-motorized vehicle. \\
One-cut-rich-or-poor & C5 & Having a barber shave a line on the scalp, risking cuts. \\
Stay wealthy & C6 & Wasting food and drinks to signal status. \\
Traveling $>$ studying & C6 & Promoting anti-education views and encouraging school dropout. \\
Betel nut & C8 & Addictive substance linked to oral cancer. \\
Cigarette brands (e.g., Zhonghua) & C8 & Cigarette brands disguised as cultural or brand names. \\
Correction fluid & C8 & Electronic cigarettes designed to resemble correction fluid bottles. \\
\bottomrule
\end{tabularx}
\end{table}

\subsection{Noise Injection}
\label{subsec:noise_injection}

We observed that noise injection was used as a circumvention strategy across multiple PHTV categories. In borderline videos involving minors—especially those classified as C2 (Child Sexual Exploitation Imagery)—creators applied visual effects such as animated face filters or digital overlays to obscure the minors’ faces. Similarly, in C8 videos depicting minors smoking, creators employed extensive facial effects to partially conceal cigarettes, reducing the recognition rate of automated visual classifiers while maintaining visibility for human viewers.

Beyond visual obfuscation, some high-risk videos incorporated misleading metadata to create a false sense of legitimacy. For instance, dangerous stunts or illicit activities were occasionally tagged with labels such as “Legal Channel,” framing harmful content as educational or cautionary.

Textual noise was also prevalent: video titles, captions, or hashtags frequently included ostensibly benign or “positive” statements—such as “I’m worried my child might go astr astray” (accompanied by promotions of e-cigarettes to minors), “All professions are worthwhile” (used alongside messages encouraging school dropout), or “Dressed normally, with no improper intentions” (paired with potentially pornographic imagery). These phrases appear designed to dilute harmful signals in textual moderation pipelines.

Overall, 23.36\% of PHTV in our dataset employed at least one form of noise injection providing a conservative lower-bound estimate of the prevalence of this evasion tactic in adolescent-facing recommendation feeds.

\section{Illustrative Examples of PHTV Categories}
\label{appendix:phtv_examples}

Below we present representative examples for each of the eight PHTV categories, illustrating how harmful content manifests in practice on short-video platforms.

\begin{itemize}
    \item \textbf{C1: Aberrant Social Media Challenges}  
    Including challenges such as imitating a dog eating something that fell on the ground, using socks to filter bubble tea, inserting food into body openings (e.g., nostrils).

    \item \textbf{C2: Child Sexual Exploitation Imagery}  
    Features minors in sexualized contexts, such as exposing body parts (nudity, slim build), wearing suggestive clothing, using harmless language (e.g., “Normal dressing style with no negative influence”), or performing suggestive dances.

    \item \textbf{C3: Glorification of Youth Violence}  
     Including fighting, bullying (“the consequence of coming to school with nothing”), armed conflicts, verbal violence, or insulting others.

    \item \textbf{C4: Intentional Self-Injury \& Extreme Body Modifications}  
    Including encompasses self-harming behaviors (cutting, burning), DIY piercings, tongue studs, tattooing.

    \item \textbf{C5: Life-Threatening Risk-Taking Behaviors}  
    Involves high-risk activities with significant potential for death or severe injury, such as human pyramids, fire-breathing, wheelie riding, dangerous haircutting, connecting alcohol lamps, making homemade explosives, dangerous driving, electric bike modifications, smashing a foot with a heavy object, or weaponized stationery modifications.

    \item \textbf{C6: Maladaptive Behavioral Influence}  
     Including dropping out of school, hating education or teachers, confronting parents, running away from home, destroying phones, or wasting food (“maintain wealth”).

    \item \textbf{C7: Predatory Recruitment of Minors}  
    Involves adults approaching minors through financial inducement or emotional manipulation, commonly asking “What grade are you in?” accompanied by a money transfer screenshot, or explicitly offering payment for companionship.

    \item \textbf{C8: Promotion of Unhealthy Lifestyles}  
    Including smoking (including eating cigarette butts), glamorizing e-cigarette sales (“afraid the child will go down the wrong path”), drinking, chewing betel nuts, resisting menstruation (“to hell with my period”), or extreme dieting.
\end{itemize}

Due to ethical considerations regarding reverse-image-search and re-identification risks, we do not include visual examples of PHTVs. Instead, we provide textual descriptions for each category below, drawn from our annotation protocol.

\section{Youth Mode Measurement Details}
\label{appendix:youth_mode}
To evaluate the effectiveness of platform-provided Youth Mode, we activated this feature on synthetic accounts on both Douyin and Kwai, following official in-app procedures. Under Youth Mode, we collected 5,863 videos from Douyin and 2,688 videos from Kwai in October 2025 using passive crawling (no interaction, no search). All videos were classified by our PHTV Arbiter and manually verified. \textbf{Zero} PHTVs were found in either dataset, indicating that Youth Mode—when enabled—effectively blocks all categories of harmful content defined in our taxonomy. This confirms the technical robustness of the filter, though its real-world impact remains limited due to low adoption among actual teens (30–41\%).

\section{In-Context Learning}
\label{appendix:ICL}
Formally, given a test input $x_{\text{test}}$, the model $\mathcal{M}$ predicts the label $y_{\text{test}}$ based on a prompt $P$. The prompt $P$ is constructed as the concatenation of the task instruction $t$, a set of demonstrations $\mathcal{D}$ (examples with labels), and the test input $x_{\text{test}}$:
\[
P = [t;\, \mathcal{D};\, x_{\text{test}}].
\]

The model predicts the label $y_{\text{test}}$ by maximizing the conditional probability over the label space $\mathcal{Y}$:
\[
y_{\text{test}} = \operatorname*{arg\,max}_{y \in \mathcal{Y}} \mathcal{M}(y \mid P),
\]
where $\mathcal{M}(y \mid P)$ represents the probability assigned by the model to label $y$ given the prompt $P$.

\section{Parameter-Efficient Fine-Tuning}
\label{appendix:PEFT}

\subject{Parameter-Efficient Fine-Tuning (PEFT)} Fine-tuning large language models (LLMs) can be computationally expensive due to their massive parameter count. Parameter-Efficient Fine-Tuning (PEFT) techniques aim to address this challenge by optimizing only a small subset of the model's parameters, or by introducing additional lightweight parameters, while keeping the majority of the pre-trained model frozen. This significantly reduces the computational and memory overhead compared to full fine-tuning, making it more practical for resource-constrained environments.

Formally, let $\mathcal{M}_{\theta}$ represent a pre-trained model with parameters $\theta$. In standard fine-tuning, all parameters $\theta$ are updated to minimize a task-specific loss $\mathcal{L}$. In contrast, PEFT modifies the model by introducing a small set of trainable parameters $\Delta \theta$, while keeping the original parameters $\theta$ fixed. The modified model can be expressed as $\mathcal{M}_{\theta + \Delta \theta}$. The optimization problem then becomes:
\[
\Delta \theta^* = \operatorname*{arg\,min}_{\Delta \theta} \mathbb{E}_{(x, y) \sim \mathcal{D}} \mathcal{L}(\mathcal{M}_{\theta + \Delta \theta}(x), y),
\]
where $\mathcal{D}$ is the training dataset, and $(x, y)$ are input-output pairs.

One prominent PEFT technique is Low-Rank Adaptation (LoRA)~\cite{hu2022lora}. LoRA assumes that the updates to the model's weights can be approximated using low-rank matrices. Specifically, for a weight matrix $W \in \mathbb{R}^{m \times n}$ in the model, LoRA introduces two trainable matrices $A \in \mathbb{R}^{m \times r}$ and $B \in \mathbb{R}^{r \times n}$, where $r \ll \min(m, n)$, and the weight update is parameterized as:
\[
\Delta W = A B.
\]
The modified weight matrix becomes:
\[
W' = W + \Delta W = W + A B.
\]
By constraining $r$ to be small, the number of trainable parameters is significantly reduced, while still allowing the model to adapt effectively to the target task.

LoRA offers several advantages that make it a compelling choice for fine-tuning large models. First, it is highly efficient. By training only the low-rank matrices $A$ and $B$, the memory and computational overhead are significantly reduced, making it suitable for resource-constrained environments. Second, it provides modularity. The low-rank updates can be stored separately, allowing for seamless switching between tasks without altering the base model. Finally, LoRA is scalable. It can be applied across multiple layers of the model, further enhancing its flexibility and adaptability to diverse tasks.

\end{document}